\documentclass{article}

\usepackage[final,nonatbib]{neurips_2022}

\usepackage[utf8]{inputenc} % allow utf-8 input
\usepackage[T1]{fontenc}    % use 8-bit T1 fonts
\usepackage{hyperref}       % hyperlinks
\usepackage{url}            % simple URL typesetting
\usepackage{booktabs}       % professional-quality tables
\usepackage{amsfonts}       % blackboard math symbols
\usepackage{nicefrac}       % compact symbols for 1/2, etc.
\usepackage{microtype}      % microtypography
\usepackage{xcolor}         % colors
\usepackage[square,numbers,sort&compress]{natbib} % need

\usepackage{graphicx}
\usepackage{subfigure}
\usepackage{booktabs} %For tables
\usepackage{amsmath, amsthm, amsfonts, amssymb}

\usepackage{algorithmic, algorithm}
\newtheorem{prop}{Proposition}
\newtheorem{theorem}{Theorem}

\newtheorem*{remark}{Remark}

\DeclareMathOperator{\diff}{d}
\usepackage{verbatim}
\usepackage{mathtools}
\usepackage{bbm}
\usepackage{authblk}
\newcommand{\beginsupplement}{
	\setcounter{table}{0}
	\renewcommand{\thetable}{S\arabic{table}}
	\setcounter{figure}{0}
	\renewcommand{\thefigure}{S\arabic{figure}}
	\setcounter{equation}{0}
	\renewcommand{\theequation}{S\arabic{equation}}}

\title{Biologically-plausible backpropagation through arbitrary timespans via local neuromodulators}

\author[1,2,3,*]{Yuhan Helena Liu}
\author[2,4]{Stephen Smith}
\author[1,2,3]{Stefan Mihalas}
\author[1,2,3]{Eric Shea-Brown}
\author[2,*]{Uygar S\"umb\"ul}
\affil[1]{Department of Applied Mathematics, University of Washington, Seattle, WA, USA}
\affil[2]{Allen Institute for Brain Science, 615 Westlake Ave N, Seattle WA, USA}
\affil[3]{Computational Neuroscience Center, University of Washington, Seattle, WA, USA}
\affil[4]{Department of Molecular and Cellular Physiology, Stanford University, Stanford CA, USA}
\affil[*]{Correspondence: hyliu24@uw.edu, uygars@alleninstitute.org}

\begin{document}
	
	\maketitle

	\begin{abstract}
		The spectacular successes of recurrent neural network models where key parameters are adjusted via backpropagation-based gradient descent have inspired much thought as to how biological neuronal networks might solve the corresponding synaptic credit assignment problem~\cite{lillicrap2020backpropagation,richards2019deep,Lillicrap2019}.
		There is so far little agreement, however, as to how biological networks could implement the necessary backpropagation through time, given widely recognized constraints of biological synaptic network signaling architectures. Here, we propose that extra-synaptic diffusion of local neuromodulators such as neuropeptides may afford an effective mode of backpropagation lying within the bounds of biological plausibility.
		Going beyond existing temporal truncation-based gradient approximations~\cite{Murray2019,Bellec2020,liu2021pnas}, our approximate gradient-based update rule, ModProp, propagates credit information through arbitrary time steps.
		ModProp suggests that modulatory signals can act on receiving cells by convolving their eligibility traces via causal, time-invariant and synapse-type-specific filter taps. Our mathematical analysis of ModProp learning, together with simulation results on benchmark temporal tasks, demonstrate the advantage of ModProp over existing biologically-plausible temporal credit assignment rules. These results suggest a potential neuronal mechanism for signaling credit information related to recurrent interactions over a longer time horizon. Finally, we derive an in-silico implementation of ModProp that could serve as a low-complexity and causal alternative to backpropagation through time. 
	\end{abstract}
	
	\section{Introduction}
	
	Recurrent connectivity is a hallmark of neuronal circuits. While this feature enables rich and flexible computation, %synaptic plasticity
	mechanisms enabling efficient task learning in large circuits remain a central problem in neuroscience and artificial intelligence research. Fundamentally, the problem stems from the fact that potentially all history of all neurons, including synaptically far away ones, can affect neuronal activity and contribute to task output. Motivated by the success of gradient descent learning, several %successful
	biological learning models %tried to
	approximate the exact gradient in recurrent neural networks using known biological processes and gain insights into computational principles of how the brain might learn~\cite{Murray2019,Bellec2020,liu2021pnas}. 
	
	By ignoring dependencies beyond a few recurrent steps --- thus severely truncating the gradient computational graph --- these existing models succeed in representing their approximate gradient-based update rule as a combination of terms that resemble known synaptic physiological processes: "eligibility trace", which maintains a fading memory of coincidental activation of presynaptic and postsynaptic neurons~\cite{Magee2020,Gerstner2018,Yagishita2014,Cassenaer2012,Sanhueza2013}, combined with a third ``modulatory'' factor ---  top-down learning or ``reward’’ signals~\cite{dayan2001theoretical,schultz2015neuronal,rubin2021credit,brzosko2019neuromodulation}, for which dopamine is a prominent candidate~\cite{schultz2016dopamine}. Despite the impressive performance of such approximations on a variety of tasks, truncating relevant credit information results in a significant performance gap compared to algorithms using exact gradient information; backpropagation-through-time (BPTT) and real time recurrent learning (RTRL)~\cite{Murray2019,Bellec2020,liu2021pnas,Marschall2019}.
	
	How might neural circuits account for long-term recurrent interactions to assign credit (or blame) to neuronal firing that happened arbitrary steps before the presentation of reward?  Given the rich repertoire of dynamical and signaling elements in the brain, one avenue could be to examine biological processes that have been underexplored in existing models. Dopamine --- whose cellular actions are exerted by activation of G protein-coupled receptors (GPCRs), which can greatly impact STDP --- is not the only neuromodulator involved in learning~\cite{cassel2020role,bazzari2019neuromodulators,Borbely2013,Hamilton2013}. More recently, transcriptomic studies have uncovered strong evidence for many other neuromodulatory pathways throughout the brain that also act via the activation of GPCRs, leading to similar downstream actions as dopamine~\cite{tasic2018, Smith2019}. This suggests that, similar to dopamine, they could also play a role in shaping synaptic credit assignment. A conspicuous family within these pathways is neuropeptide signaling because peptidergic genes are densely and abundantly expressed in the forebrains of divergent species, including the human, in a cell-type-specific fashion~\cite{smith2021transcriptomic}, suggesting widespread interaction between synaptic and peptidergic modulatory networks for synaptic credit assignment~\cite{Smith2020, liu2021pnas}. Moreover, intra-cortical expression allows neuropeptides to potentially carry information \textbf{local} to the cortical network, cell type specificity enables sculpted signals for different recipient cells, and their diffusive nature could enable communication between neurons that are not synaptically connected. Finally, peptidergic signals have timescales much longer than the time scales for axonal propagation of action potentials or synaptic delays~\cite{VandenPol2012}. Taken together, these properties make cell-type-specific local neuromodulation seem promising for propagating credit signal over multiple recurrent steps. Developing explicit computational principles of how these local modulatory elements could propagate credit signal over arbitrary recurrent steps could advance our understanding of biological learning and may inspire more efficient low-complexity bio-inspired learning rules.
	
	Motivated by the question above as well as shortcomings of gradient approximations based on severe temporal truncations, we investigate how biological credit signals could be propagated through arbitrary recurrent steps via widespread cell-type-specific neuropeptidergic signaling~\cite{Smith2019,Smith2020}. While Ref.~\cite{liu2021pnas} recently introduced a framework exploiting properties of neuropeptidergic signaling for temporal credit assignment, similar to~\cite{Murray2019} and~\cite{Bellec2020}, their approach performs severe temporal truncation of the error gradient and does not consider credit propagation beyond disynaptic connections. Our main contributions are summarized as follows:
	\begin{itemize}
		\item We derive a theory that provides mechanism and intuition for the effectiveness of synapse-type-specific modulatory backpropagation (through time) weights (Theorem~\ref{thm:decE}). 
		\item We develop a model predicting how modulatory signaling could be the basis for biological backpropagation through \textbf{arbitrary} time spans (Figures~\ref{fig:schematics1_mdgl++} and~\ref{fig:schematics2_mdgl++}).  Unlike temporal truncation-based approximations~\cite{Murray2019,Bellec2020,liu2021pnas}, our model enables each neuron to receive filtered (rather than precise) credit signals regarding its contribution to the outcome via neuromodulation. 
		\item We demonstrate the effectiveness of modulatory signaling via synapse-type-specific, rather than synapse-specific, modulatory weights on learning tasks that involve temporal credit assignment (Figure~\ref{fig:learningCurves_MDGL++Wab}). In particular, we demonstrate an \textbf{online} learning setting where weights are updated causally and in real-time (Figure~\ref{fig:learningCurvesMDGL++_online}).
		\item We also derive a low-complexity in silico implementation of our algorithm suitable for \textbf{online learning} (Proposition~\ref{prop:online}). 
	\end{itemize}
	
	\section{Related works}
	
	%\subsection{Neuromodulatory factors in synaptic plasticity}
	
	{\bf Neuromodulatory factors in synaptic plasticity:} One of the most fundamental learning rules, Hebbian plasticity, attributes lasting changes in synaptic strength and memory formation to correlations of spike timing between particular presynaptic and postsynaptic neurons~\cite{dan2006spike,song2000competitive}. However, multiple experimental and theoretical investigations now indicate that the Hebbian rule alone is insufficient. First, there have been numerous suggestions that some persistent ``eligibility trace’’ must exist to bridge the temporal gap between correlated firings at millisecond timescales and behavioral timescales lasting seconds~\cite{Magee2020,Gerstner2018,Cassenaer2012,Yagishita2014,lehmann2019one,Suvrathan2019}. Moreover, impacts of correlated spike timing must be augmented by one or more additional modulatory factors to steer weight updates toward desired outcome~\cite{Farries2007,schultz2016dopamine,pawlak2010modstdp,brzosko2019neuromodulation,Roelfsema2018,Magee2020,lillicrap2020backpropagation,payeur2020burst,aljadeff2019cortical,Fremaux2015,Gerstner2018,florian2007reinforcement,pogodin2020kernelized}. This is commonly known as learning with three factors. A prominent candidate for such a modulatory factor is dopamine~\cite{schultz2015neuronal}. Dopamine influences receiving neurons via activation of G protein-coupled receptors (GPCRs), which can regulate membrane excitability and key parameters of synaptic plasticity rules. 
	
	Besides dopamine, recent transcriptomic evidence has uncovered genes encoding a plethora of other neuromodulatory pathways throughout the brain, including neuropeptide signaling genes that are abundantly expressed in the forebrains of tetrapods, including the human~\cite{Smith2019, smith2021transcriptomic}. Like dopamine, their cellular actions are exerted by the activation of G protein-coupled receptors (GPCRs), which can persistently modulate Hebbian synaptic plasticity~\cite{brzosko2019neuromodulation,Tritsch2012,marder2012neuromodulation}. This suggests that they too, could play a role in shaping cortical learning and synaptic credit assignment. Moreover, nearly all neurons express one or more neuropeptide signaling gene~\cite{Smith2019}, which suggests a dense interplay between synaptic and peptidergic modulatory networks to shape synaptic credit assignment~\cite{Smith2020}. 
	
	%\subsection{Approximate gradient descent learning}
	
	{\bf Approximate gradient descent learning:} Standard algorithms for gradient descent learning in recurrent neural networks (RNNs), real time recurrent learning (RTRL) and backpropagation through time (BPTT), are not biologically-plausible~\cite{Richards2019} and have vast computational and storage demands~\cite{Williams1995}. However, multiple studies have shown that learning algorithms that approximate the gradient, while mitigating some of the problems of exact gradient computation, can lead to satisfactory learning outcomes~\cite{Richards2019,linsley2020,lillicrap2016FA}. In feedforward networks, plenty of biologically-plausible learning rules have been proposed and demonstrated impressive performance that rival backpropagation on many different tasks ~\cite{Roelfsema2018,Richards2019,rubin2021credit,lillicrap2016FA,payeur2020burst,pozzi2018biologically,sacramento2018dendritic,laborieux2021scaling,amit2019deep,millidge2020activation}. 
	
	For efficient online learning in RNNs, approximations to RTRL have been proposed~\cite{Marschall2019,Mujika2018,Tallec2018,Roth2019,Murray2019,Menick2020,zenke2020spike}. For instance, a recent influential study~\cite{Bellec2020} conceived how the three-factor learning framework could approximate the gradient. Among the biologically-plausible proposals~\cite{Murray2019,Bellec2020,liu2021pnas}, approximations have mainly been based on temporal truncations of the gradient computation graph and because of that, their ability to learn dependencies across arbitrary recurrent steps has been limited. Lastly, non-truncation-based approximations have been proposed outside of bio-plausible research~\cite{Mujika2018,Tallec2018}. For further discussions on related algorithms, please refer to Appendix~\ref{scn:related_algs}.
	
	\section{Results}
	
	\begin{figure}[h!]
		\centering
		\includegraphics[width=\textwidth]{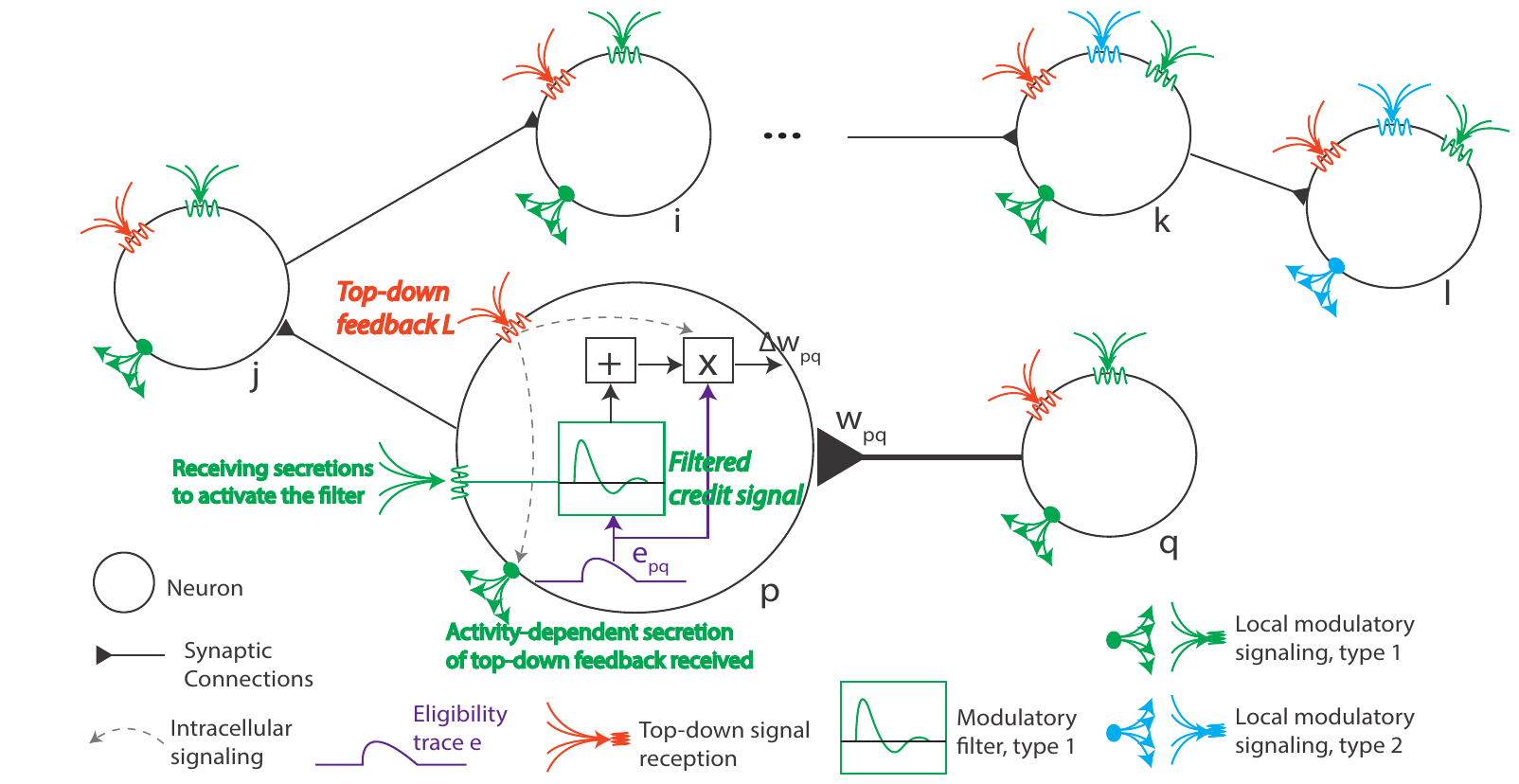}
		\caption{\textbf{Biologically-plausible temporal credit assignment via modulatory and synaptic message passing.} %local modulatory networks on top of synaptic networks.}
		In addition to established biological learning ingredients (eligibility traces and a top-down learning signal~\cite{fremaux2016neuromodulated,Gerstner2018}), synapse-type-specific local modulatory networks may also be involved in weight updates~\cite{Smith2020}. Our learning rule, ModProp, conceives the action of participating modulators on receiving cells as a convolution of the eligibility trace with causal, time-invariant, cell-type-specific filters. Each circle represents a neuron and the synaptic weight of interest is $W_{pq}$; we illustrate the cellular processes of postsynaptic neuron $p$. Our derivation (Supp. Section~\ref{scn:methods} and Theorem~\ref{thm:decE}) predicts that the modulatory signal each neuron receives can represent a filtered credit signal regarding how its past firings (arbitrary steps back) contribute to the task outcome. 
	}
	\label{fig:schematics1_mdgl++}
\end{figure}

\begin{figure}[h!]
	\centering
	\includegraphics[width=\textwidth]{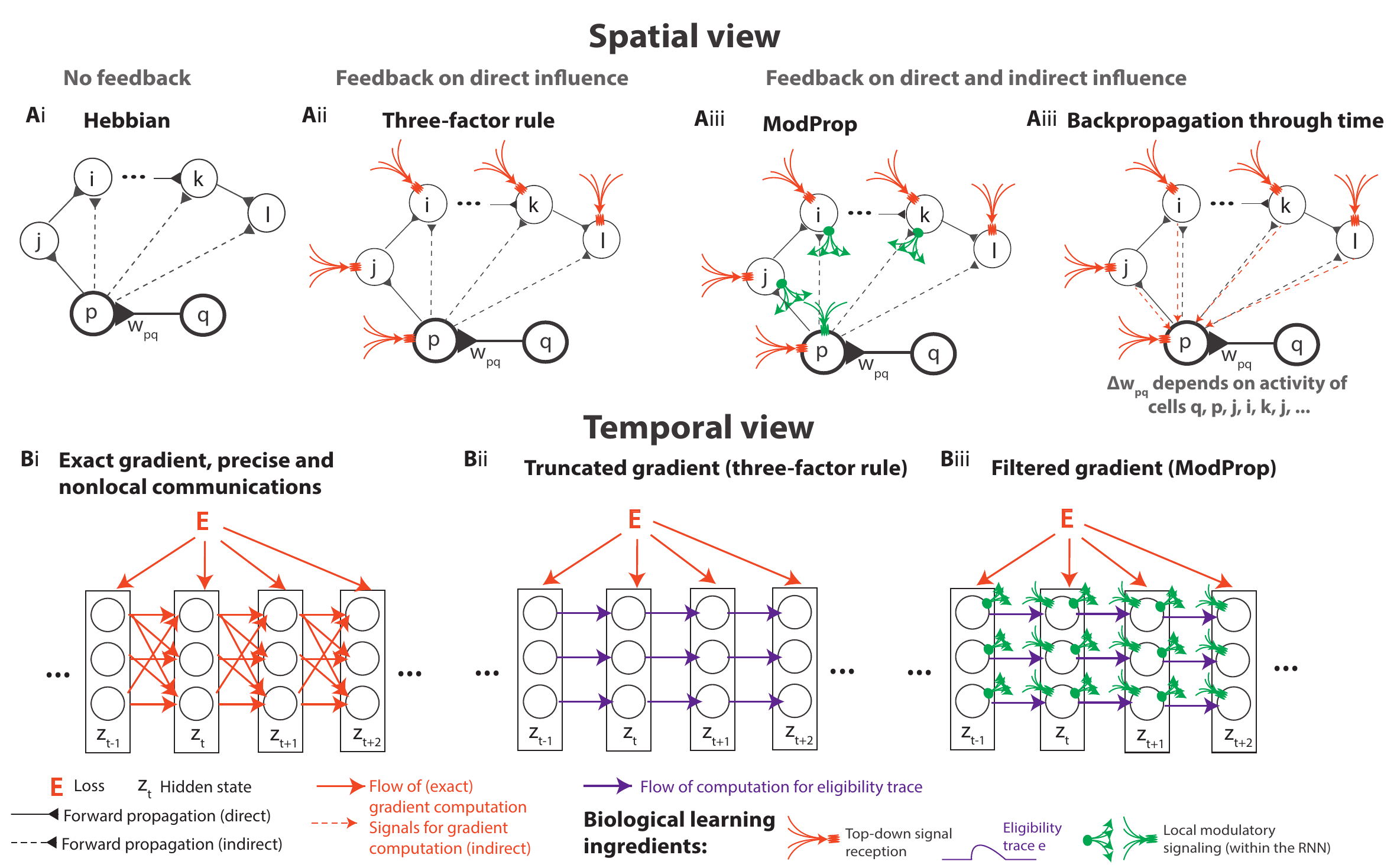}
	\caption{\textbf{Local modulatory signaling for gradient estimation.} A) A spatial view of learning rules for updating weight $W_{pq}$. (i) Hebbian learning, where weight update depends only on pre-/post-synaptic activities. (ii) Three-factor learning~\cite{fremaux2016neuromodulated, Murray2019, Bellec2020}, which updates weights using additional top-down learning signals, severely truncates the exact gradient. (iii) ModProp also accounts for (filtered) distant feedback information delivered through synapse-type-specific neuromodulation; (iv) BPTT computes the exact gradient: weight update involves nonlocal information, i.e. activities of indirectly connected units. B) A temporal view. Bi) BPTT propagates the precise intercellular dependencies in an acausal manner. Bii) Three-factor learning rule neglects all the intercellular dependencies in the temporal propagation of the credit signal. Biii) ModProp \textbf{approximates} such spatiotemporal dependencies through local neuromodulatory signals (Eq.~\ref{eq_ModProp}). ModProp approximates the exact gradient by assuming similar connectivity among cells of the same type, and filtering the indirect effects on loss from neurons that are potentially many synapses away. Figure~\ref{fig:ModProp_approx} provides a summary of approximations made by ModProp. }
	\label{fig:schematics2_mdgl++}
\end{figure}

\subsection{Learning rule overview}

Biological plausibility is a guiding principle in developing our model. Hence, the model choices are based either on established constraints of neurobiology such as the locality of synaptic transmission, causality, and Dale’s Law, or on emerging evidence from large-scale datasets, such as the cell-type-specificity of certain neuromodulators~\cite{Smith2019} or the hierarchical organization of cell types~\cite{tasic2018}. We consider a discrete-time rate-based RNN similar to the form in~\cite{ehrlich2021psychrnn} with observable states, i.e. firing rates, as $z_{t}$ at time $t$, and the corresponding internal states as $s_t$. $W$ denotes the recurrent weight matrix with $(pq)^\text{th}$ entry $W_{pq}$ representing the synaptic connection strength between presynaptic neuron $q$ and postsynaptic neuron $p$. See Supp. Section~\ref{scn:methods} for the neuron model. 

\textbf{Gradient descent learning in RNNs}: RNNs are typically trained by gradient descent learning on the task error (or negative reward) $E$. However, the two equivalent factorizations of error gradient in RNNs, BPTT and RTRL, both involve nonlocal information that is inaccessible to neural circuits. This is due to recurrent connectivity: a synaptic weight $W_{pq}$ can affect loss $E$ through many other neurons in the other network in addition to its pre- and postsynaptic neurons. To see this more concretely, the following is the exact gradient at time $t$ using RTRL factorization, on which we base our approximation for online learning:
%\begin{align}
%\frac{\diff E}{\diff W_{pq}} |_t  &=\sum_{j}\frac{\partial E}{\partial s_{j,t}} \frac{\diff s_{j,t}}{\diff W_{pq}} \label{eqn:sum_tj0} \\
%    \frac{\diff s_{j,t}}{\diff W_{pq}} &= \frac{\partial s_{j,t}}{\partial W_{pq}} + \sum_l \frac{\partial s_{j,t}}{\partial s_{l,t-1}} \frac{\diff s_{l,t-1}}{\diff W_{pq}} \cr
%    &= \frac{\partial s_{j,t}}{\partial W_{pq}} + \frac{\partial s_{j,t}}{\partial s_{j,t-1}} \frac{\diff s_{j,t-1}}{\diff W_{pq}} + 
%    {\underbrace{\textstyle \sum_{l\neq j} W_{jl} \frac{\partial z_{l,t-1}}{\partial s_{l,t-1}} \frac{\diff s_{l,t-1}}{\diff W_{pq}} }_{\mathclap{ \text{\normalsize depends on all weights $W_{jl}$}}}}, \label{eqn:s_triple0}
%\end{align}
\begin{align}
	\frac{\diff E}{\diff W_{pq}} |_t  &=\sum_{j}\frac{\partial E}{\partial s_{j,t}} \frac{\diff s_{j,t}}{\diff W_{pq}} \label{eqn:sum_tj0} \\
	\frac{\diff s_{j,t}}{\diff W_{pq}} &= \frac{\partial s_{j,t}}{\partial W_{pq}} + \sum_l \frac{\partial s_{j,t}}{\partial s_{l,t-1}} \frac{\diff s_{l,t-1}}{\diff W_{pq}} = \frac{\partial s_{j,t}}{\partial W_{pq}} + \frac{\partial s_{j,t}}{\partial s_{j,t-1}} \frac{\diff s_{j,t-1}}{\diff W_{pq}} + 
	{\underbrace{\textstyle \sum_{l\neq j} W_{jl} \frac{\partial z_{l,t-1}}{\partial s_{l,t-1}} \frac{\diff s_{l,t-1}}{\diff W_{pq}} }_{\mathclap{ \text{\normalsize depends on all weights $W_{jl}$}}}}, \label{eqn:s_triple0}
\end{align}
where $\partial$ denotes direct dependency and $\diff$ accounts for all (direct and indirect) dependencies, following the notation in~\cite{Bellec2020}. (See Appendix~\ref{scn:gd_rnn} for further details on the notation.) While $\frac{\partial E}{\partial s_{j,t}}$, which considers only the direct contribution of the internal state of neuron $j$ at time $t$ to the loss, is easy to compute, the factor $\frac{\diff s_{j,t}}{\diff W_{pq}}$ is a memory trace of all inter-cellular dependencies and requires $O(N^3)$ memory and $O(N^4)$ computations. This makes RTRL expensive to implement for large networks. Moreover, the last factor $\frac{\diff s_{j,t}}{\diff W_{pq}}$ poses a serious problem for biological plausibility: the nonlocal terms in Eq.~\ref{eqn:s_triple0} requires knowledge of all other weights in the network to update the weight $W_{pq}$. Existing biologically-plausible solutions to this problem apply severe truncations: references~\cite{Murray2019} and~\cite{Bellec2020} completely ignore the third nonlocal term (Figure~\ref{fig:schematics2_mdgl++}), whereas reference~\cite{liu2021pnas} restores terms within one recurrent step through local modulatory signaling but truncates further terms. 

\textbf{ModProp approximation overview:} Derivation of gradient descent-based weight updates involving neuromodulatory signaling from Eq.~\ref{eqn:sum_tj0} suggests two approximations to Eq.~\ref{eqn:sum_tj} and~\ref{eqn:s_triple0} to move beyond severe truncations of the exact gradient while retaining biological plausibility. \textbf{Approximation 1} replaces the activation derivative with a constant:
\begin{equation}
	\frac{\partial z_{j,t}}{\partial s_{j,t}} \approx \mu \quad \text{(approximation 1)}, \label{eqn:approx1}
\end{equation}
where $\mu$ represents the average activity of neurons in a ReLU network. This approximation assumes stationarity in neuron activity and uncorrelatedness of such activity with a small subset of synaptic weights, as explained in derivations leading up to Appendix Eq.~\ref{eqn:WH_nMDGL} and Eq.~\ref{eqn:mu_filter} (Methods in Appendix~\ref{scn:methods}). While neuronal activity and synaptic weights are not necessarily uncorrelated, considering that a single neuron may have thousands of synaptic partners, the activity of the neuron or its time derivative is typically weakly correlated to any one synaptic weight. This also indicates that such approximation might work better for large networks with many neurons and synaptic partners for each neuron, as is the case for biological neural networks. We take advantage of this phenomenon in our model and ignore these weak correlations. This approximation enables the filter taps (Eq.~\ref{eq_ModProp}) to be \textbf{time-invariant}, a property likely required for biological plausibility. We also define $S$, the arbitrary number of credit propagation steps, and it will become clear later that this corresponds to the number of modulatory filter taps. With this, the estimated gradient becomes: 
\begin{align}
	\frac{\diff E}{\diff W_{pq}}|_t 
	& \approx  \frac{\partial E}{\partial z_{p,t}} e_{pq,t} + \sum_j \frac{\partial E}{\partial s_{j,t}} \sum_{s=1}^{S} (W^s)_{jp}\, \mu^{s-1} \, e_{pq,t-s} \label{eq_ModProp_0}
\end{align}
where $e_{pq,t} \approx \frac{d z_{p,t}}{ d W_{pq}}$ ($e_{pq,t}$ is defined precisely in Appendix Eq.~\ref{eqn:etrace}) can be interpreted as a persistent Hebbian ``eligibility trace’’~\cite{Cassenaer2012, Yagishita2014, Gerstner2018} that keeps a fading memory of past coincidental pre- and postsynaptic activity~\cite{Bellec2020}. Here, $(W^s)_{jp}$ represents the $(jp)^\text{th}$ entry of $W^s$, $W$ raised to the power of $s$, and $(W^1)_{jp}=W_{jp}$. Details of the derivation can be found in Appendix~\ref{scn:methods}. 

It is important to note that \textbf{Approximation 1} is applied only to the nonlocal gradient terms, i.e. exact activation derivative is used during the computation of the eligibility trace that is local to the pre- and postsynaptic neurons. Also, we treat $\mu$ as a hyperparameter in our simulations and explain how this is tuned in Appendix~\ref{scn:sim_details}. One should note that tuning the hyperparameter $\mu$ properly is important, because the estimated gradient can explode when $\mu$ is too large and ModProp approaches the three-factor rule when $\mu$ is too small. Future work involves improving ModProp with an adaptive $\mu$ for better numerical stability and accuracy.

\textbf{Approximation 2} replaces \textbf{synapse-specific} feedback weights with \textbf{type-specific} weights:
\begin{equation}
	(W^s)_{jp} \approx (W^s)_{\alpha \beta} \quad \text{(approximation 2)}. \label{eqn:approx2}
\end{equation}
Here, cell $j$ belongs to type $\alpha$, cell $p$ is of type $\beta$ and $C$ denotes the set of cell types. (e.g., $W_{\alpha \beta}=\mathbb{E}_{j\in\alpha,p\in\beta}[W_{jp}]$, $\alpha,\beta\in C$.) This approximation is due to the type-specific nature of modulatory channels~\cite{Smith2019}. We call these modulatory weights synapse-type-specific (as opposed to cell-type-specific) to emphasize the connectivity-based grouping. Details of how these modulatory weights were obtained can be found in Appendix~\ref{scn:methods}. 

\textbf{ModProp: filtering credit signals via local neuromodulation:} Substituting \textbf{Approximation 1} and \textbf{Approximation 2} into Eq.~\ref{eqn:sum_tj} and~\ref{eqn:s_triple0} leads to the ModProp update:  
\begin{align}
	\left. \Delta W_{pq} \right|_{ModProp}
	&\propto L_p \times e_{pq} + \Bigg( \sum_{\alpha \in C} \bigg( \sum_{j\in \alpha} \text{L}_j \text{ activity}_j \bigg)\times F_{\alpha \beta}\Bigg) * \text{e}_{pq} , \cr
	F_{\alpha \beta, s} &= \mu^{s-1} (W^s)_{\alpha \beta}, \label{eq_ModProp}
\end{align}
where $L$ and $e$ denote top-down learning signal and eligibility trace, respectively. We postulate that $F_{\alpha \beta}$ represents type-specific filter taps of GPCRs expressed by cells of type $\beta$ to precursors secreted by cells of type $\alpha$; $*$ is the convolution operation with $S$ as the number of filter taps. Note that the matrix powers $(W^s)_{\alpha \beta}$ appearing in the values of different filter taps $F_{\alpha \beta, s}$ could be genetically pre-determined as part of the cell identity and optimized over evolutionary time scales. (Appendix Figure~\ref{fig:LearningCurveMDGL++_FixWab} shows successful learning using fixed modulatory weights.) Details of a biological interpretation of Eq.~\ref{eq_ModProp} can be found in Appendix~\ref{scn:cost_plausible}. Briefly, the secretion of top-down (TD) learning signals can selectively activate a biochemical process at the post-synaptic neuron, which can then act as a temporal filter on the eligibility trace.

Observing that neurons of the same type demonstrate consistent properties across a range of features (e.g. connectivity, physiology, gene expression)~\cite{Smith2019,gouwens2019classification}, we use two cell types with consistent wiring, neuromodulation, and type of synaptic action (excitatory/inhibitory) in our relatively simple models.

In summary, we propose a synaptic weight update rule, where the eligibility trace is compounded not only with top-down learning signals -- as in modern biologically-plausible learning rules~\cite{Magee2020,Gerstner2018} -- but also with local modulatory pathways through convolution (Figure~\ref{fig:schematics1_mdgl++}). This modulatory mechanism allows the propagation of credit signals through an arbitrary number of recurrent steps. 

\subsection{Properties of ModProp}

\begin{remark} \label{rmk:cost}
	The biologically plausible implementation (Eq.~\ref{eq_ModProp}) of ModProp complexity scales as $O(SN^2)$ per time step $t$. 
\end{remark}
Here, $N$ and $S$ denote the number of recurrent units and the number of filter taps. As seen in Eq.~\ref{eq_ModProp_0}, the number of filter taps corresponds to the number of recurrent steps for which the credit information is propagated. We also present an alternative implementation with potentially lower cost later in Proposition~\ref{prop:online}. Details for cost analysis and biological interpretation can be found in Appendix~\ref{scn:cost_plausible}. 

We next show through Theorem~\ref{thm:decE} that learning with synapse-type-specific weights leads to loss reduction at every step on average. The mechanistic intuition behind Theorem~\ref{thm:decE} is that, in the presence of a statistical connection between synaptic weights (forward path) and modulatory weights (feedback path), the modulatory feedback signal each neuron receives can be a good estimate of how its activity contributes to the overall task error, which can be used for effective tuning of incoming synaptic weights to reduce the error. Figure~\ref{fig:plotAlignMDGL++} compares the angle with the exact gradient across different learning algorithms to demonstrate that the direction of the approximate gradient computed by ModProp is similar to (aligned with) the direction of the exact gradient, thereby reducing the loss.

In the following, we define the ``residual weights'' away from the cell-type averages as $\epsilon_{ij}:=W_{ij}-W_{\alpha \beta}$ and $(\epsilon_s)_{ij}:=(W^s)_{ij}-(W^s)_{\alpha \beta}$, where $i \in \alpha, j \in \beta$, $\alpha, \beta \in C$. We consider these terms as stochastic so that the circuit output and the eligibility traces are also stochastic, as functions of $\epsilon_s$. We consider the cell type averages and the output connection strengths as deterministic. (Merely modifying the uncorrelatedness assumption below would extend our results to stochastic output connection strengths.) Below, $\mathbb{E}$ is short for $\mathbb{E}_\epsilon$, $\left. \Delta E \right|_{t}$ denotes the change in the task error at time $t$ and $\left. \Delta E \right|_{pq,t}$ denotes the contribution of the synapse $(pq)$ to it.
\begin{theorem} \label{thm:decE} 
	Consider linear RNNs $\mathcal{N}$, $\widehat{\mathcal{N}}$ with weight matrices $W$, $\widehat{W}$, respectively and identical architectures. Let $\widehat{W}_{ij}=W_{\alpha\beta}$, $\forall i\in\alpha, j\in\beta$. Assume a small enough learning rate such that the remainder of the first-order Taylor expansion of loss is negligible. For network $\mathcal{N}$, if $\mathbb{E}[\epsilon_{ij}]=0$, $(\epsilon_{s})_{ik}$ and $\epsilon_{kj}$ are uncorrelated for $i\neq j$, and $\epsilon_s$ and $(y_t-y_t^*)^2  e_{pq,t'}e_{pq,t-s}$ are uncorrelated for any $s,t,t'$, then $\left. \mathbb{E}[\Delta E \right|_{pq,t}] \leq 0$ and $\left. \mathbb{E}[\Delta E \right|_{t}] \leq 0$. Moreover, $\left. \mathbb{E}[\Delta E \right|_{t}] < 0$ if gradient descent is possible for network $\widehat{\mathcal{N}}$. (Proof is in Appendix~\ref{scn:proofs}.)
\end{theorem}

%\begin{proof}
%Proof is in Appendix~\ref{scn:proofs}. 
%\end{proof}
Note that the uncorrelatedness assumption above is relatively mild because any single connection strength is typically a very poor predictor of network activity, especially as the network size grows. Note also that gradient update can introduce a drift in residual weights, requiring a similar drift in cell-type averages for the strict inequality to be applicable over multiple update steps.

\subsection{Simulation results}

\begin{figure}[h!]
	\centering
	\includegraphics[width=\textwidth,trim=9 0 20 0, clip]{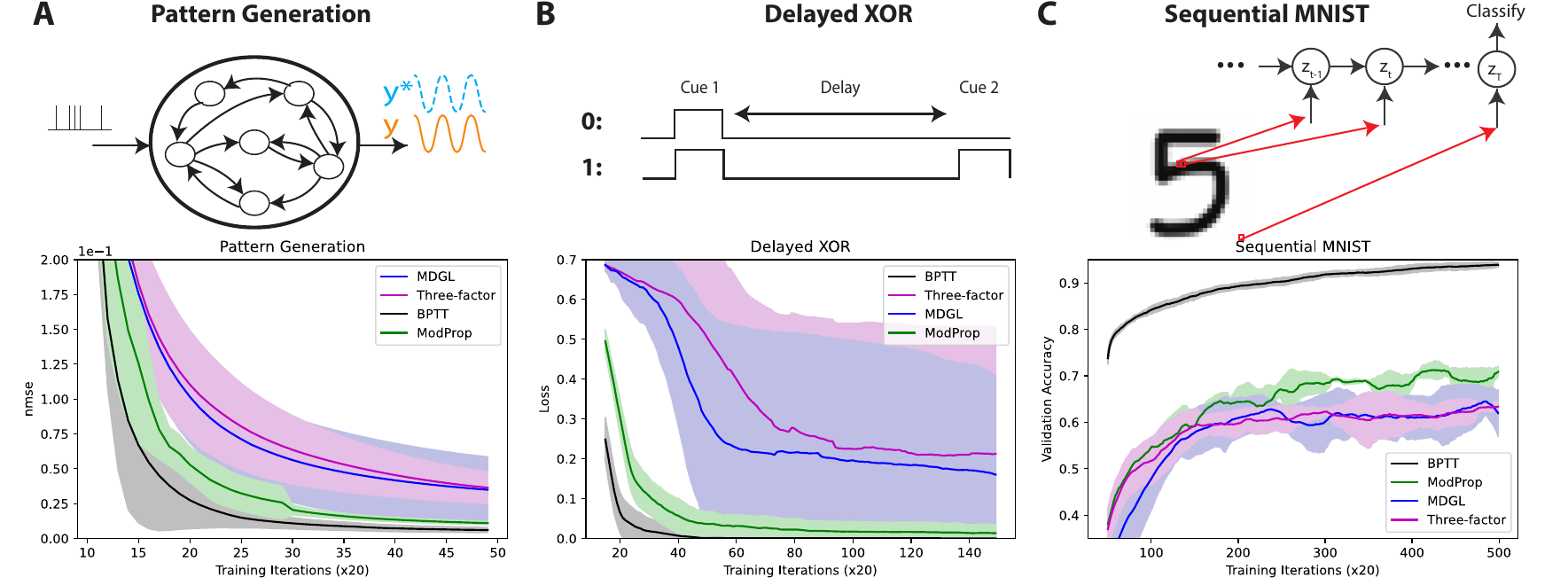}
	\caption{\textbf{Modulatory signaling of credit information on long-term recurrent interactions can improve learning outcomes.} ModProp improves the learning performance over existing bio-plausible rules. This experiment examines the performance due to Approximation 1 (Eq.~\ref{eqn:approx1}) before any cell-type approximation of modulatory weights (Eq.~\ref{eqn:approx2}). %See Fig.~\ref{fig:learningCurves_MDGL++Wab} for the effect of cell-type approximation.
		A) Learning to produce a time-resolved target output pattern. B) A delayed XOR task, where the network determines if two cue alternatives --- the presence or absence of input represented by $1$ or $0$ --- match or mismatch after a delay, requiring memory via recurrent activity. c) Pixel-by-pixel MNIST task~\cite{lecun1998mnist}. Note that this task is unlikely to be solved effectively by humans. (See text.) Consistent with the original MDGL paper~\cite{liu2021pnas}, we also find that MDGL confers little advantage over the three-factor rule (e-prop) under dense connectivity. Solid lines/shaded regions: mean/standard deviation of loss curves across five runs.}
	\label{fig:learningCurves_MDGL++}
\end{figure}

\begin{figure}[h!]
	\centering
	\includegraphics[width=0.8\textwidth,trim=9 0 30 0, clip]{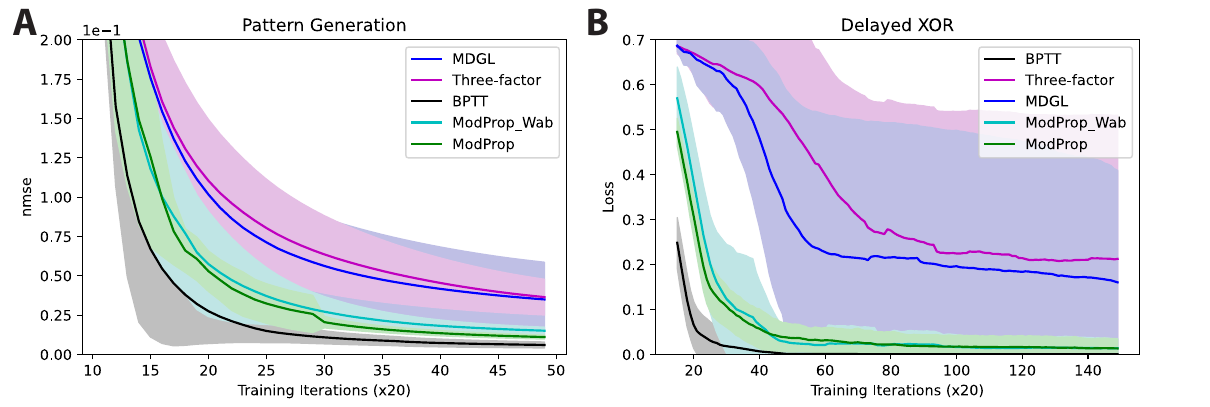}
	\caption{\textbf{Efficient learning with type-specific modulatory weights.} In addition to Approximation 1 studied in Figure~\ref{fig:learningCurves_MDGL++}, this figure investigates the effect of Approximation 2 (labeled as ModProp\_Wab), which uses type-specific, rather than synapse-specific feedback weights for signaling credit information (Eq.~\ref{eqn:approx2}). Here, ModProp\_Wab uses only two modulatory types mapped to the two main cell classes. The cell type approximation does not result in any significant performance degradation in A) the pattern generation task and B) the delayed XOR task. This analysis is not done for the sequential MNIST task, where neurons were not divided into E and I types. Solid lines/shaded regions: mean/standard deviation of loss curves across five runs.}
	\label{fig:learningCurves_MDGL++Wab}
\end{figure}

To test the ModProp formulation, we study its performance in well-known tasks involving temporal processing: pattern generation, delayed XOR, and sequential MNIST. We compare the learning performance of ModProp with the state-of-the-art biologically-plausible learning rules (MDGL~\cite{liu2021pnas} and e-prop~\cite{Bellec2020} (labeled as ``three-factor''), as well as BPTT to provide a lower bound on task error. Note, BPTT and RTRL both compute the exact gradient, so they should be identical in terms of performance. Here, we chose BPTT due to computational efficiency. Consistent with~\cite{liu2021pnas}, we also found MDGL to confer little advantage over e-prop when all neurons are connected to the readout  (Figure~\ref{fig:learningCurves_MDGL++}), so we focus our analysis on that case. 

We first study pattern generation with RNNs, where the aim is to produce a one-dimensional target output, generated from the sum of five sinusoids, given a fixed Gaussian input realization. We change the target output and the random input along with initial weights for different training runs, and show the learning curve in Figure~\ref{fig:learningCurves_MDGL++}A across seven such runs. By using a densely connected network, we observe that MDGL confers little advantage over the three-factor rule (e-prop), as reported in the original paper~\cite{liu2021pnas}. Moreover, communicating longer indirect effects, despite being filtered, leads to improved learning outcomes, as demonstrated by the superior performance of ModProp over MDGL. 

Next, to study how RNNs learn to process discrete cues that impact delayed outcomes, we consider a delayed XOR task: two cue alternatives, $1$ or $0$, are encoded by the presence/absence of input. The network is trained to remember the first cue and learn to compare it with the second cue delivered at a later time to determine if the two cues match. Figure~\ref{fig:learningCurves_MDGL++}B illustrates the learning curves for this task. We observe the same general conclusion as the previous task. Some learning curves have a standard deviation, which indicates that the network struggled to learn the task with these rules for some seeds; based on the standard deviation of the curves, it seems possible for ModProp to perform similarly as three-factor for some seeds, but the focus here is to examine performance across many runs. In Appendix Figure~\ref{fig:learningCurves_varDly}, we also examined the learning performance of the same set of learning rules for a longer delay period. Interestingly, the performance of ModProp approaches that of BPTT for this task and the previous one, further closing the gap between artificial network training and biological learning mechanisms in performing credit assignment over a long period.

Finally, we study the pixel-by-pixel MNIST~\cite{lecun1998mnist} task, which is a popular machine learning benchmark. Although it is not a task that the brain would solve well (i.e., humans would struggle to predict the digit with only one pixel presented at a time), we investigate it to test the limits of spatiotemporally filtered credit signals for tasks that demand temporally precise input integration. Figure~\ref{fig:learningCurves_MDGL++}C illustrates the learning curves for this task. While the performance ordering of learning rules is still the same as in previous tasks, we observe a wider gap between ModProp and BPTT. Since the time-invariant filter approximation (Eq~\ref{eqn:approx1}) restricts the spatiotemporal resolution of the credit signal, this is in line with our expectation that ModProp will struggle with tasks that demand highly precise spatiotemporal integration, such as the pixel-by-pixel MNIST task that even the brain would struggle to solve.  

As a proof-of-concept study, we initially focused our analysis on the approximation performance by imposing time-invariant filter taps (Eq.~\ref{eqn:approx1}). Next, in Figure~\ref{fig:learningCurves_MDGL++Wab}, we investigate the learning performance using type-specific rather than synapse-specific feedback modulatory weights (Eq.~\ref{eqn:approx2}). These type-specific weights were calculated using weight averages as in~\cite{liu2021pnas}. Later, we repeat these simulations using fixed random modulatory weights in Appendix Figure~\ref{fig:LearningCurveMDGL++_FixWab}; we also demonstrate the superior performance of ModProp with fixed random modulatory weights for the ``copy task''~\cite{mujika2018approximating} in Appendix Figure~\ref{fig:learningCurves_copy}. Little performance degradation is observed for only two modulatory types mapped onto the two main cell classes, indicating the effectiveness of cell-type discretization. Other than the sequential MNIST task, cells are divided into two main cell classes --- with $80\%$ of the cells being excitatory (E) and $20\%$ being inhibitory (I) --- and obeyed connection sign constraints. 

\begin{figure}[h!]
	\centering
	\includegraphics[width=0.8\textwidth,trim=9 0 30 0, clip]{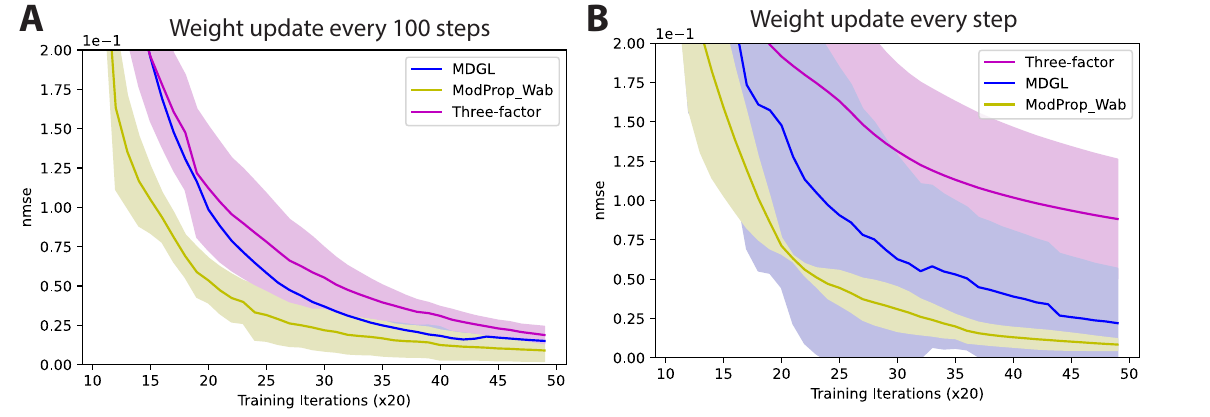}
	\caption{\textbf{Superior performance of ModProp in an online learning setting}. We investigate an online learning version of the pattern generation task, where weights are updated either A) every 100 time steps or B) every single step. Here, ModProp uses the efficient online learning implementation derived in Proposition~\ref{prop:online}. Plotting conventions follow those of previous figures. }
	\label{fig:learningCurvesMDGL++_online}
\end{figure}

% We omit comparisons with the exact gradient update because online learning requires the computation of the RTRL factorization, whose complexity is prohibitive. 

Lastly, we demonstrate the advantage of our weight update rule in an online learning setting of the pattern generation task, where weights are updated in real time (Figure~\ref{fig:learningCurvesMDGL++_online}). For that, we also derive a cost and storage efficient in silico implementation of ModProp in Proposition~\ref{prop:online}. 

\begin{prop} \label{prop:online}
	ModProp has an online in-silico (not necessarily biologically-plausible) implementation with $O(CN^2)$ storage and $O(C^2N^2)$ computational complexity, where $C$ is the number of cell types. (Proof is in Appendix~\ref{scn:cost_implausible}.)
\end{prop}
%\begin{proof}
%Proof is in Appendix~\ref{scn:cost_implausible}. 
%\end{proof}

\section{Discussion}
A central question in the study of biological and artificial intelligence is how the temporal credit assignment problem is solved in neuronal circuits~\cite{Lillicrap2019}.  Motivated by recent genomic evidence on the widespread presence of local modulatory networks~\cite{Smith2019,smith2021transcriptomic}, we demonstrated how such signaling could complement Hebbian learning and global neuromodulation (e.g., via dopamine) to achieve biologically plausible propagation of the credit signal through arbitrary time steps. Going beyond the scalar feedback provided by global top-down modulation, our study proposes how detailed vector feedback information~\cite{lillicrap2020backpropagation} can be delivered in neuronal circuits. Instead of the severe temporal truncations of the gradient information proposed by the state-of-the-art~\cite{Murray2019,Bellec2020,liu2021pnas}, ModProp offers a framework where the full temporal feedback signal can be received, albeit via low-pass filtering at the post-synaptic neuron due to specificity at the level of neuronal types, and not individual neurons. Moreover, predictions generated by ModProp on the role of a family of signaling molecules (e.g. neuropeptides) could potentially be tested experimentally: physiology of multiple individual cells can be monitored in modern neurobiology experiments~\cite{Smith2020,melzer2020bombesin}. Blocking certain peptidergic receptors of the neurons that are involved with learning a task and comparing the performance to that without blocking can provide a test for the role of peptidergic communication.

\begin{figure}[h!]
	\centering
	\includegraphics[width=0.8\textwidth]{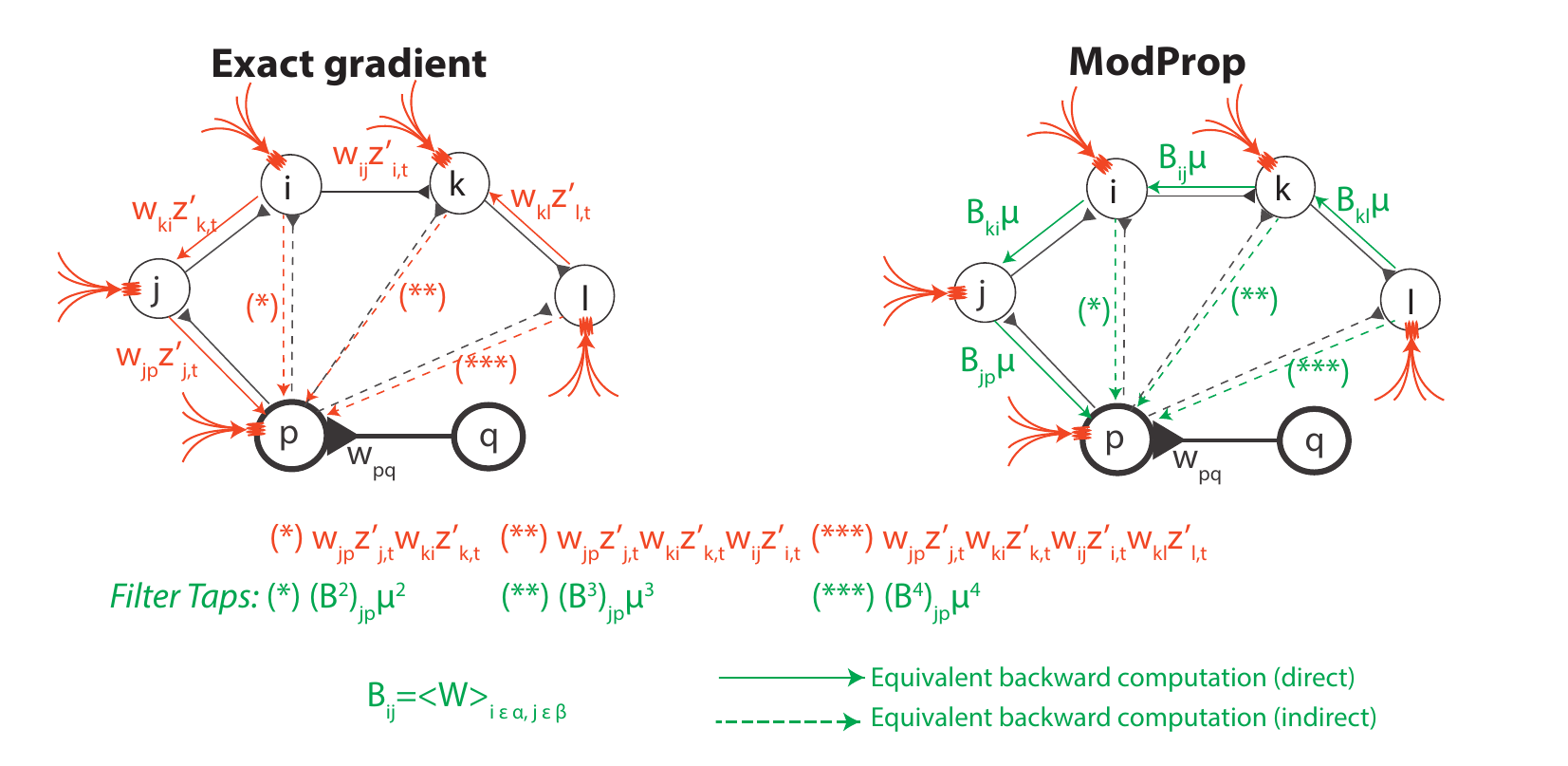}
	\caption{\textbf{ModProp brings two approximations to the nonlocal gradient terms. } First, ModProp uses type-specific feedback weights (modulatory weights), rather than cell-specific feedback weights (Eq.~\ref{eqn:approx1}). Second, ModProp approximates the activation derivative (Eq.~\ref{eqn:approx2}). Similar to feedback alignment for feedforward networks~\cite{lillicrap2016FA}, different weights are used during the backward pass than the forward pass due to the well-known weight transport problem. This, however, won't be adequate for RNN settings. On top of the type-specific feedback weights approximation, time-invariant activation derivative approximation was also applied for time-invariant filtering (as explained in the texts surrounding Eq.~\ref{eq_ModProp_0} and Eq.~\ref{eq_ModProp}). Any unspecified symbols in the illustration were defined in Figure~\ref{fig:schematics2_mdgl++}. }
	\label{fig:ModProp_approx}
\end{figure}

While feedback alignment~\cite{lillicrap2016FA} addresses the weight transport problem in feedforward networks, it is not clear in RNNs which biological pathways would implement \textbf{temporal} feedback. Our model suggests that such pathways could come from synapse-type-specific local neuromodulation. In addition to improved performance compared to the state-of-the-art across all of our simulations, our theoretical and experimental results show that ModProp can be implemented efficiently for online learning. Figure~\ref{fig:ModProp_approx} briefly mentions some of the connections of ModProp to feedback alignment. Together, these findings suggest that synapse-type-specific local modulatory signaling could be a neural mechanism for propagating credit information over more than just a few recurrent steps.

Among the many future directions, a natural extension could be to investigate the performance of ModProp across a broader range of tasks. This could include situations where the assumptions in deriving the rule (e.g., stationarity of activity) are severely violated. This might improve the approximations introduced here. Similarly, we observed a significant gap between ModProp and BPTT for the pixel-by-pixel MNIST task that demands precise temporal integration of input (Figure~\ref{fig:learningCurves_MDGL++}C), but we discussed earlier that this is a challenging task for the brain. Additionally, this study focuses on dense networks with ReLU activation; future directions include investigating two biologically relevant paradigms: sparse connectivity and a diverse set of activation functions (spike-based in particular). Although ModProp can be applied in theory to temporal credit assignment over an arbitrary duration, the presented learning rule accounts for dependencies at every single step (as for BPTT). This means that, similar to BPTT, it is ill-suited for very long-term credit assignment~\cite{yang2021next}. An interesting line of research attends to the issue of extracting relevant, rather than full, memories~\cite{yang2021next,lillicrap2019backpropagation}. Investigating how our approximations could potentially be combined with memory sparsification techniques to perform very long-term biologically-plausible credit assignment can be fruitful~\cite{yang2021next}. 

While our paper advances the basic science of learning and we do not foresee immediate societal impacts of our work, its benefits to both neuroscience and deep learning research could have long-term (positive or negative) societal impacts. It is hard to overstate the philosophical implications of understanding how the brain works (and, in particular, learns). Moreover, such understanding could guide us toward curing certain diseases of the brain. Nevertheless, in the same vein, such future tools could also enable abuse if left unregulated. On the machine learning side, our method can be considered a low-cost alternative to the ubiquitous BPTT algorithm. In this sense, our algorithm or a future method that builds on it can make tasks that are currently approachable by just a few wealthy entities available to many more practitioners. On the flip side, as with many other capable machine learning tools, such entities are also capable of utilizing low-cost learning to conquer even harder tasks, which is potentially a reason for concern. Finally, data-driven tools will ultimately reflect the various biases in their training data and our method is no exception. 

\section*{Acknowledgements} {\label{sec:acknowledgements}}
We thank the Allen Institute founder, Paul G Allen, for his vision, encouragement and support. YHL is supported by the NSERC PGS-D program. This work was facilitated through the use of the UW Hyak supercomputer system.

\bibliographystyle{unsrt}
\bibliography{ref_MDGL}

%%%%%%%%%%%%%%%%%%%%%%%%%%%%%%%%%%%%%%%%%%%%%%%%%%%%%%%%%%%%

\section*{Checklist}

%%% BEGIN INSTRUCTIONS %%%
The checklist follows the references.  Please
read the checklist guidelines carefully for information on how to answer these
questions.  For each question, change the default \answerTODO{} to \answerYes{},
\answerNo{}, or \answerNA{}.  You are strongly encouraged to include a {\bf
	justification to your answer}, either by referencing the appropriate section of
your paper or providing a brief inline description.  For example:
\begin{itemize}
	\item Did you include the license to the code and datasets? \answerYes{See Section~\ref{gen_inst}.}
	\item Did you include the license to the code and datasets? \answerNo{The code and the data are proprietary.}
	\item Did you include the license to the code and datasets? \answerNA{}
\end{itemize}
Please do not modify the questions and only use the provided macros for your
answers.  Note that the Checklist section does not count towards the page
limit.  In your paper, please delete this instructions block and only keep the
Checklist section heading above along with the questions/answers below.
%%% END INSTRUCTIONS %%%

\begin{enumerate}

	\item For all authors...
	\begin{enumerate}
		\item Do the main claims made in the abstract and introduction accurately reflect the paper's contributions and scope?
		\answerYes{We have supported our key contribution points (see Introduction) with simulations and mathematical proofs.}
		\item Did you describe the limitations of your work?
		\answerYes{Please see discussion on future direction in the Discussion section.}
		\item Did you discuss any potential negative societal impacts of your work?
		\answerYes{Please see the Discussion section.}
		\item Have you read the ethics review guidelines and ensured that your paper conforms to them?
		\answerYes{We have carefully read the ethics review guidelines and confirm that this
			paper conforms to them.}
	\end{enumerate}

	\item If you are including theoretical results...
	\begin{enumerate}
		\item Did you state the full set of assumptions of all theoretical results?
		\answerYes{We have made sure that all assumptions are precisely stated in the Theorem statement. Details for where these assumptions come into the proof can be found in Appendix~\ref{scn:proofs}.}
		\item Did you include complete proofs of all theoretical results?
		\answerYes{We have provided complete and detailed proofs in Appendix~\ref{scn:proofs}.}
	\end{enumerate}

	\item If you ran experiments...
	\begin{enumerate}
		\item Did you include the code, data, and instructions needed to reproduce the main experimental results (either in the supplemental material or as a URL)?
		\answerYes{Anonymized code link is provided in Appendix~\ref{scn:sim_details}.}
		\item Did you specify all the training details (e.g., data splits, hyperparameters, how they were chosen)?
		\answerYes{Please see Appendix~\ref{scn:sim_details}.}
		\item Did you report error bars (e.g., with respect to the random seed after running experiments multiple times)?
		\answerYes{We ran experiments across random seeds and reported uncertainty for all applicable figures. We explained the plotting convention in the caption.}
		\item Did you include the total amount of compute and the type of resources used (e.g., type of GPUs, internal cluster, or cloud provider)?
		\answerYes{Please see Appendix~\ref{scn:sim_details}.}
	\end{enumerate}

	\item If you are using existing assets (e.g., code, data, models) or curating/releasing new assets...
	\begin{enumerate}
		\item If your work uses existing assets, did you cite the creators?
		\answerYes{Please see Appendix~\ref{scn:sim_details}.}
		\item Did you mention the license of the assets?
		\answerYes{Please see Appendix~\ref{scn:sim_details}.}
		\item Did you include any new assets either in the supplemental material or as a URL?
		\answerNo{}
		\item Did you discuss whether and how consent was obtained from people whose data you're using/curating?
		\answerNA{}
		\item Did you discuss whether the data you are using/curating contains personally identifiable information or offensive content?
		\answerNA{}
	\end{enumerate}

	\item If you used crowdsourcing or conducted research with human subjects...
	\begin{enumerate}
		\item Did you include the full text of instructions given to participants and screenshots, if applicable?
		\answerNA{}
		\item Did you describe any potential participant risks, with links to Institutional Review Board (IRB) approvals, if applicable?
		\answerNA{}
		\item Did you include the estimated hourly wage paid to participants and the total amount spent on participant compensation?
		\answerNA{}
	\end{enumerate}

\end{enumerate}

%%%%%%%%%%%%%%%%%%%%%%%%%%%%%%%%%%%%%%%%%%%%%%%%%%%%%%%%%%%%

\newpage 

\appendix

\beginsupplement
\section{Methods} \label{scn:methods}

\subsection{Network Model}

We consider a discrete-time implementation of a rate-based recurrent neural network (RNN) similar to the form in~\cite{ehrlich2021psychrnn}. We denote the observable states, i.e. firing rates, as $z_{t}$ at time $t$, and the corresponding internal states as $s_t$. The dynamics of those states are governed by 
\begin{align}
	s_{j,t+1} &= \eta\, s_{j,t} + (1-\eta) \left(\sum_{l\neq j} W_{jl}\, z_{l,t} + \sum_p W_{jm}^\text{IN}\, x_{m,t+1} \right)  \cr
	z_{j,t} &= ReLU(s_{j,t}), \label{eq_rate1}
\end{align}
where $\eta=e^{-dt/\tau_m}$ denotes the leak factor for simulation time step $dt$ and membrane time constant $\tau_m$, $W_{lj}$ denotes the weight of the synaptic connection from neuron $j$ to $l$, $W_{jm}^\text{IN}$ denotes the strength of the connection between the $m^{th}$ external input and neuron $j$ and $x_t$ denotes the external input at time $t$. Threshold adaptation is not used here in order to focus on capacity of the temporal credit propagation mechanism. We focused on ReLU activation due to its wide adoption in both deep learning and computational neuroscience communities; as discussed, we leave extension to other activation functions (spike-based in particular) for future work.

The readout $y$ is a linear transformation of the hidden state
\begin{equation}
	y_{k,t}=\sum_j W_{kj}^\text{OUT} z_{j,t} + b_k^\text{OUT},
	\label{eq_outputNeuron}
\end{equation}
where $W_{kj}^\text{OUT}$ denotes the strength of the connection from neuron $j$ to output neuron $k$, $b_k^\text{OUT}$ denotes the bias of the $k$-th output neuron.

We quantify how well the network output matches the desired target using loss function $E$:
\begin{equation}
	E = \begin{cases}
		\frac{1}{2} \sum_{k,t} (y^*_{k,t} - y_{k,t})^2, \text{ for regression tasks} \cr
		-\sum_{k,t} \pi^*_{k,t} log \pi_{k,t}, \text{ for classification tasks} 
	\end{cases} \label{eqn:loss_fcn}
\end{equation}
where $y^*_{k,t}$ is the time-dependent target, $\pi^*_{k,t}$ is the one-hot encoded target and $\pi_{k,t} = \text{softmax}_k(y_{1,t}, \dots, y_{N_{OUT},t}) = \exp(y_{k,t})/\sum_{k'} \exp(y_{k',t})$ is the predicted category probability. 

\subsection{Gradient descent learning in RNNs} \label{scn:gd_rnn}

\textbf{Notation for Derivatives}: There are two types of computational dependencies in RNNs; direct and indirect dependencies. We distinguish direct dependencies versus all dependencies (including indirect ones) using partial derivatives ($\partial$) versus total derivatives ($d$), respectively. 

Without loss of generality, consider a function $f(x,y)$, where $y$ itself may depend on $x$. The partial derivative $\partial$ of $f$ at $x_0$ considers $y$ as a constant, and evaluates as $\frac{\partial f (x,y)}{\partial x} |_{x_0,y(x_0)}$; i.e. the derivative calculation only considers how $x$ directly affects $f$. The total derivative $\diff$, on the other hand, may not treat $y$ as a constant and evaluates as $\frac{d f(x,y)}{d x} = \frac{\partial f (x,y)}{\partial x} |_{x_0,y(x_0)} + \frac{\partial f (x,y)}{\partial y} |_{x_0,y(x_0)} \frac{\partial y}{\partial x} |_{x_0}$; i.e. the derivative calculation also takes into account how $x$ can indirectly affect $f$ through $y$.

As an example in our network, variable $W_{pq}$ can impact state $s_{p,t}$ directly through Eq.~\ref{eq_rate1}, i.e. $\frac{\partial s_{p,t}}{W_{pq}} = (1-\eta) z_{q,t-1}$. On the other hand, $W_{pq}$ can also impact $s_{p,t}$ indirectly through other cells in the network: i.e. the dependence of $s_{p,t}$ on $W_{pq}$ and all $s_{j,t'}$ ($t'<t$, $j \in \{1, ..., N\}$) affected by $W_{pq}$ are taken into account for the derivative calculation, which leads to the recursive equation in Eq.~\ref{eqn:s_triple}.

\textbf{Exact gradient computation and locality issue}:
In gradient descent learning, all weight parameters (input weights $W^{IN}$, recurrent weights $W$ and output weights $W^{OUT}$) are adjusted iteratively according to the error gradient. This error gradient can be calculated with classical machine learning algorithms, backpropagation through time (BPTT) and real time recurrent learning (RTRL)~\cite{Williams1995}, which uses different factorization but yield equivalent results. However, the BPTT factorization depend on future activity, which poses an obstacle for online learning and biological plausibility. Our learning rule derivation follows the RTRL factorization because it is causal. 

RTRL factors the error gradient across time and space as
\begin{align}
	\frac{\diff E}{\diff W_{pq}} |_t  &=\sum_{j}\frac{\partial E}{\partial s_{j,t}} \frac{\diff s_{j,t}}{\diff W_{pq}} \label{eqn:sum_tj} \\
	\frac{\diff z_{j,t}}{\diff W_{pq}} &= h_{j,t} \frac{\diff s_{j,t}}{\diff W_{pq}}, \text{where } h_{j,t}:=\frac{\partial z_{j,t}}{\partial s_{j,t}}    \label{eqn:dzdw} \\
	\frac{\partial s_{j,t}}{\partial W_{pq}} &= \delta_{jp} (1-\eta) z_{q,t-1} \\
	\frac{\partial s_{j,t}}{\partial s_{l,t-1}} &= \begin{cases}
		\eta, & j=l \\
		\frac{\partial s_{j,t}}{\partial z_{l,t-1}} \frac{\partial z_{l,t-1}}{\partial s_{l,t-1}} = (1-\eta) W_{jl} h_{l,t-1}, & j\neq l \end{cases} \\
	\frac{\diff s_{j,t}}{\diff W_{pq}} &= \frac{\partial s_{j,t}}{\partial W_{pq}} + \sum_l \frac{\partial s_{j,t}}{\partial s_{l,t-1}} \frac{\diff s_{l,t-1}}{\diff W_{pq}} \cr 
	&= \frac{\partial s_{j,t}}{\partial W_{pq}} + \eta \frac{\diff s_{j,t-1}}{\diff W_{pq}} + (1-\eta)
	{\underbrace{\textstyle \sum_{l\neq j} W_{jl} \frac{\partial z_{l,t-1}}{\partial s_{l,t-1}} \frac{\diff s_{l,t-1}}{\diff W_{pq}} }_{\mathclap{ \text{\normalsize depends on all weights $W_{jl}$}}}}. \label{eqn:s_triple}
\end{align}
following the derivative notation explained above. The factor $\frac{\partial E}{\partial z_{j,t}}$ in Eq.~\ref{eqn:sum_tj} can be interpreted as the top-down learning signal, which is defined as $L_{j,t} := \sum_k W^{OUT}_{kj} (y_{k,t} - y^*_{k,t})$ for regression tasks~\cite{Bellec2020}. It is straightforward to compute. However, the triple tensor $\frac{\diff s_{j,t}}{\diff W_{pq}}$ requires $O(N^3)$ memory and $O(N^4)$ computation costs. It keeps track of all the paths that $z_{j,t}$ can affect $W_{pq}$ (for every $j, p, q$). Moreover, it poses a significant challenge to biological plausibility: updating each weight $W_{pq}$ requires knowing all other weights $W_{jl}$ (for every $j$ and $l$) in the network, and that information should be inaccessible to neural circuits. 

To address this, references~\cite{Murray2019} and~\cite{Bellec2020} dropped the problematic terms so that the updates to weight $W_{pq}$ would only depend on pre- and post-synaptic activity, and applied this truncation to train rate- and spike-based networks, respectively. However, such truncation results in limited performance. 

\subsection{Derivation of ModProp}

We ask how intercellular neuromodulation might communicate the expensive spatiotemporal dependency in the factor $\frac{\diff s_{j,t}}{\diff W_{pq}}$. Along with the interpretation of $L_{j,t} := \frac{\partial E}{\partial z_{j,t}}$ as top-down learning signal, let $e_{pq,t}$ denote the eligibility trace of coincidental activation between presynaptic cell $q$ and postsynaptic cell $p$~\cite{Bellec2020}. The following derivation leads to our learning rule. The leak factor is omitted in the derivation below ($\eta=0$) for readability, and we substitute Eq.~\ref{eqn:s_triple} into Eq.~\ref{eqn:sum_tj} and repeatedly expand the expensive $\frac{\diff s}{\diff W}$ factor using Eq.~\ref{eqn:s_triple}:
\begin{align}
	\frac{\diff E}{\diff W_{pq}}|_t &= \sum_j \frac{\partial E}{\partial z_{j,t}} h_{j,t} \left(\delta_{jp}z_{q,t-1} + \sum_l W_{jl} h_{l,t-1} \frac{\diff s_{l,t-1}}{\diff W_{pq}} \right) \\
	&= \frac{\partial E}{\partial z_{p,t}} h_{p,t} z_{q,t-1} + \sum_j \frac{\partial E}{\partial z_{j,t}} h_{j,t} \sum_l W_{jl} h_{l,t-1} \frac{\diff s_{l,t-1}}{\diff W_{pq}} \\
	&= \frac{\partial E}{\partial z_{p,t}} e_{pq,t} + \sum_j \frac{\partial E}{\partial z_{j,t}} h_{j,t} \sum_l W_{jl} h_{l,t-1} \left(\delta_{lp}z_{q,t-2} + \sum_k W_{lk} h_{k,t-2} \frac{\diff s_{k,t-2}}{\diff W_{pq}} \right) \\
	&= \ldots \nonumber
\end{align}
\begin{align}
	\frac{\diff E}{\diff W_{pq}}|_t &= \frac{\partial E}{\partial z_{p,t}} e_{pq,t} + \left(\sum_j \frac{\partial E}{\partial z_{j,t}} h_{j,t} W_{jp}\right) e_{pq,t-1} + \cr &
	\sum_j \frac{\partial E}{\partial z_{j,t}} h_{j,t} \sum_{s=2}^{S} 
	\sum_{i_1, \ldots,i_{s-1}} W_{ji_1} W_{i_1 i_2} \ldots W_{i_{s-1} p} h_{i_1,t-1}\ldots h_{i_{s-1},t-s+1} e_{pq,t-s}  \label{eqn:pre_nMDGL} \\
	& \overset{(a)}{\approx} \frac{\partial E}{\partial z_{p,t}} e_{pq,t} + \left(\sum_j \frac{\partial E}{\partial z_{j,t}} h_{j,t} W_{jp}\right) e_{pq,t-1} \cr 
	&\quad + \sum_j \frac{\partial E}{\partial z_{j,t}} h_{j,t} \sum_{s=2}^{S} (W^s)_{jp}\, e_{pq,t-s} \frac{1}{N^{s-1}} \sum_{i_1, \ldots,i_{s-1}}  h_{i_1,t-1} h_{i_2,t-2} \ldots h_{i_{s-1},t-s+1}   \\
	& =  \frac{\partial E}{\partial z_{p,t}} e_{pq,t} + \sum_j \frac{\partial E}{\partial z_{j,t}} h_{j,t} \sum_{s=1}^{S} (W^s)_{jp}\, H(t,s) \, e_{pq,t-s} \label{eqn:WH_nMDGL} ,
\end{align}
where $H(t,1)=1$, $H(t,s)=\frac{1}{N^{s-1}} \sum_{i_1, \ldots,i_{s-1}} h_{i_1,t-1} h_{i_2,t-2} \ldots h_{i_{s-1},t-s+1}$ for $s = 1, \dots, S$ and $S$, as explained later, is the number of filter taps. Again, we neglected the leak factor in the derivation for readability but included in the actual simulations. The only approximation step above, $(a)$, is made by using a point estimate assuming that the $W$ and $h$ chains are uncorrelated and the central limit theorem applies. We note that in a linear network, all the activation derivatives $h$ would be $1$, making the approximation exact. 

We expand on our explanation for step $(a)$ approximation. We define a $W$-chain (of length $l$) as
\begin{equation}
	\prod_{\phi=1}^{l} W_{i_\phi i_{\phi+1}},
\end{equation} 

for any indices $i_1,\ldots,i_{l+1} \in \{1, ..., N\}$. Similarly, we define an $h$-chain (of length $l’$) as
\begin{equation}
	\prod_{\theta=1}^{l'} h_{j_\theta, t-\theta},
\end{equation} 

for any indices $j_1,\ldots,j_{l’} \in \{1, ..., N\}$. With these definitions, we call the $W$-chain $W_{i_1,\ldots,i_{s-1}} = W_{j i_1} W_{i_1 i_2} \ldots W_{i_{s-1} p}$ and the $h$-chain $h_{i_1,\ldots,i_{s-1}} = h_{i_1, t-1} \ldots h_{i_{s-1},t-s+1}$ uncorrelated if
\begin{equation}
	\mathbb{E}_{i_1,\ldots,i_{s-1}} W_{i_1,\ldots,i_{s-1}} h_{i_1,\ldots,i_{s-1}} = \mathbb{E}_{i_1,\ldots,i_{s-1}} W_{i_1,\ldots,i_{s-1}} \mathbb{E}_{i_1,\ldots,i_{s-1}} h_{i_1,\ldots,i_{s-1}}.
\end{equation} 

Considering $W_{i_1,\ldots,i_{s-1}}$ and $h_{i_1,\ldots,i_{s-1}}$ as random i.i.d. samples indexed by $i_1,\ldots,i_{s-1}$, the central limit theorem states that
\begin{equation}
	\sum_{i_1,\ldots,i_{s-1}} W_{i_1,\ldots,i_{s-1}} h_{i_1,\ldots,i_{s-1}} \sim \mathcal{N}(N^{s-1} \mathbb{E}[W_{i_1,\ldots,i_{s-1}} h_{i_1,\ldots,i_{s-1}}], N^{s-1} \text{Var}(W_{i_1,\ldots,i_{s-1}} h_{i_1,\ldots,i_{s-1}}))
\end{equation} 
as the sum tends to infinity. Here, we simply use the i.i.d. assumption even though stronger versions of the Central Limit Theorem need weaker assumptions than i.i.d. When the $W$- and $h$-chains are uncorrelated, we take the mean of this distribution as a point estimate (note, however, the growing variance) to arrive at the following approximation:
\begin{equation}
	\sum_{i_1,\ldots,i_{s-1}} W_{i_1,\ldots,i_{s-1}} h_{i_1,\ldots,i_{s-1}} \approx N^{s-1} \mathbb{E}W_{i_1,\ldots,i_{s-1}} h_{i_1,\ldots,i_{s-1}} = N^{s-1} \mathbb{E}W_{i_1,\ldots,i_{s-1}} \mathbb{E} h_{i_1,\ldots,i_{s-1}}.
\end{equation} 

Since $\mathbb{E}W_{i_1,\ldots,i_{s-1}} = \frac{1}{N^{s-1}} (W^s)_{jp}$ (by the same application of central limit theorem as above) when $i_1,\ldots,i_{s-1}$ are distributed uniformly over valid index ranges, we conclude that 
\begin{equation}
	\sum_{i_1,\ldots,i_{s-1}} W_{i_1,\ldots,i_{s-1}} h_{i_1,\ldots,i_{s-1}} \approx (W^s)_{jp} \frac{1}{N^{s-1}} \sum_{i_1,\ldots,i_{s-1}} h_{i_1,\ldots,i_{s-1}},
\end{equation} 
where the expectation of the $h$-chain is replaced by its empirical estimate.

%%%
% In other words, the sum is related to the mean due to the law of large numbers, so $\sum_{i_1, \ldots,i_{s-1}} W_{ji_1} W_{i_1 i_2} \ldots W_{i_{s-1} p} h_{i_1,t-1}\ldots h_{i_{s-1},t-s+1} \approx N^s \mathbb{E}_{i_1, \ldots,i_{s-1}} W_{ji_1} W_{i_1 i_2} \ldots W_{i_{s-1} p} h_{i_1,t-1}\ldots h_{i_{s-1},t-s+1}$, and then we break up the expectation using the uncorrelatedness assumption $N^s \mathbb{E}_{i_1, \ldots,i_{s-1}} W_{ji_1} W_{i_1 i_2} \ldots W_{i_{s-1} p} h_{i_1,t-1}\ldots h_{i_{s-1},t-s+1} = N^s \mathbb{E}_{i_1, \ldots,i_{s-1}} W_{ji_1} W_{i_1 i_2} \ldots W_{i_{s-1} p} \mathbb{E}_{i_1, \ldots,i_{s-1}} h_{i_1,t-1}\ldots h_{i_{s-1},t-s+1}$, and then turn that back into the summation form using the law of large numbers $\frac{1}{N^s} \sum_{i_1, \ldots,i_{s-1}} W_{ji_1} W_{i_1 i_2} \ldots W_{i_{s-1} p} \sum_{i_1, \ldots,i_{s-1}} h_{i_1,t-1}\ldots h_{i_{s-1},t-s+1}$

To put this derivation in terms of biological components, we make the following further approximations. First, we link modulatory weights to type-specific GPCR efficacies, which means they are type-specific, i.e. $(W^s)_{jp} \approx (W^s)_{\alpha\beta}$, for type-indices $\alpha$ and $\beta$ in a set of possible classes $C$. Second, $H(t,s)$ should be time-invariant, i.e. $H(t,s)=H(s)$, since biological filter properties should not vary rapidly across time. 

Interestingly, we observe that $H(s)$ is an average of activation history across time and cells (Eq.~\ref{eqn:WH_nMDGL}). In particular, when the activation function is ReLU, one can think of $H(s)$ as approximating the number of activation chains with length $s$ (divided by the total number of possible chains). Thus, a crude starting point would be to assume first-order stationarity, i.e., assume the average activity level remains invariant ($\frac{1}{N} \sum_i h_{i,t} := \mu_t \approx \mu, \forall t$). Then 
\begin{align}
	H(t,s) & = \frac{1}{N^{s-1}} \sum_{i_1, \ldots,i_{s-1}} h_{i_1,t-1} \ldots h_{i_{s-1},t-s+1} \cr 
	&\overset{(a)}{=} \frac{1}{N} \sum_{i_1} h_{i_1,t-1} \frac{1}{N} \sum_{i_2} h_{i_2,t-2} \ldots \frac{1}{N} \sum_{i_{s-1}} h_{i_{s-1},t-s+1} \cr 
	&\approx \mu^{s-1}, \label{eqn:mu_filter}
\end{align}
where $\mu$ is a scalar constant and represents global average neuron activation. $(a)$ is because in the case of ReLU activation (where activation derivative $h$ is binary), total number of different activation chain combinations, $\sum_{i_1, \ldots,i_{s-1}} h_{i_1,t-1} \ldots h_{i_{s-1},t-s+1}$, would equal to the product of number of activation at each time step, $\sum_{i_1} h_{i_1,t-1} \sum_{i_2} h_{i_2,t-2} \ldots \sum_{i_{s-1}} h_{i_{s-1},t-s+1}$. This is because to choose a chain of activated neurons from all possible combinations, the number of possible indices to choose from for each step is equal to the number of activated neuron at that step. Indeed, intercellular signaling level is activity-dependent~\cite{VandenPol2012}. For implementation, $\mu$ can be treated as a hyperparameter or adapted on a separate timescale. It is also important to note that this activation derivative approximation is only applied to the term (in the equation below) additional to the e-prop term, $\frac{\partial E}{\partial z_{p,t}} e_{pq,t}$, for which the exact activation derivative is still used. 
% To understand this, consider at time 1 there are $A_1$ activated neurons, time 2 there are $A_2$ activated neurons, and time T there are $A_T$ activated neurons, then the total number of possible combinations of a chain of activated neurons is $A_1 * A_2 * … A_T*$.

By substituting these further approximations into Eq.~\ref{eqn:WH_nMDGL}, the approximated gradient becomes:
\begin{align}
	\frac{\diff E}{\diff W_{pq}}|_t 
	& \approx  \frac{\partial E}{\partial z_{p,t}} e_{pq,t} + \sum_{\alpha\in C} \left (\sum_{j\in\alpha} \frac{\partial E}{\partial z_{j,t}} h_{j,t} \right) \sum_{s=1}^{S} (W^s)_{\alpha\beta}\, \mu^{s-1} \, e_{pq,t-s}, 
\end{align}
and this leads to the ModProp update:  
\begin{align}
	\left. \Delta W_{pq} \right|_{ModProp}
	&\propto L_p \times e_{pq} + ( \sum_{\alpha \in C} \left( \sum_{j\in \alpha} \text{L}_j h_j \right)\times F_{\alpha \beta}) * \text{e}_{pq} , \cr
	F_{\alpha \beta, s} &= \mu^{s-1} (W^s)_{\alpha \beta},
	\label{eq_MDGL++}
\end{align}
where cell $j$ is of type $\alpha$, cell $p$ is of type $\beta$ and $C$ denotes the set of cell types. $L$ and $e$ denote top-down learning signal and eligibility trace, respectively. Again, activation derivative $h_j$ is closely linked to activity level of neuron $j$. $F$ represents the modulatory filter, $F_{\alpha \beta} * \text{e}_{pq} = \sum_{s=1}^S F_{\alpha \beta, s} \text{e}_{pq, t-s}$ is the convolution operation with $S$ as the number of filter taps, and scaling factor $\mu$ is a hyperparameter. For calculating the modulatory weights, the weights were calculated using matrix powers for  $s>1$. (See beginning of Theorem~\ref{thm:decE} proof.) For $s=1$, we first examined $W^{1}_{\alpha \beta} = <W^{1}_{jp}>_{j \in \alpha, p \in \beta}$ in the main text. This assumes modulatory weights and synaptic weights co-adapt throughout training and to what extent they co-adapt in neural circuits is unclear. Thus, we also set modulatory weights to fixed random type-specific values and demonstrate the resulting learning performance in Appendix Figure~\ref{fig:LearningCurveMDGL++_FixWab}. These fixed random type-specific modulatory weights were generated randomly from the distribution of averages of random initial weights. Thus, these fixed random type-specific modulatory weights would be close to the initial synaptic weight averages and could stay close depending on how much these synaptic weight averages change throughout training.  

\textbf{Eligibility trace implementation}: We here explain the implementation of eligibility trace $e_{pq,t}$:
\begin{align}
	e_{pq,t} &:= \frac{\partial z_{p,t}}{\partial s_{p,t}} \epsilon_{pq,t}, \label{eqn:etrace} \\
	\epsilon_{pq,t} &= \frac{\partial s_{p,t}}{\partial w_{pq}} + \frac{\partial s_{p,t}}{\partial s_{p,t-1}} \epsilon_{pq,t-1}, \label{eqn:evec}
\end{align}
which tracks the coincidence of postsynaptic activity $h_{p,t} = \frac{\partial z_{p,t}}{\partial s_{p,t}}$ and a low pass filtering of presynaptic activity stored in $\epsilon_{pq,t}$ ($\frac{\partial s_{p,t}}{\partial w_{pq}}=z_{q,t-1}$ and $\frac{\partial s_{p,t+1}}{\partial s_{p,t}}=\eta$ following Eq.~\ref{eq_rate1}). Reference~\cite{Bellec2020} provides a comprehensive discussion on how eligibility traces can be interpreted as derivatives.

\newpage

\section{Additional simulations}

\begin{figure}[h!]
	\centering
	\includegraphics[width=0.99\textwidth]{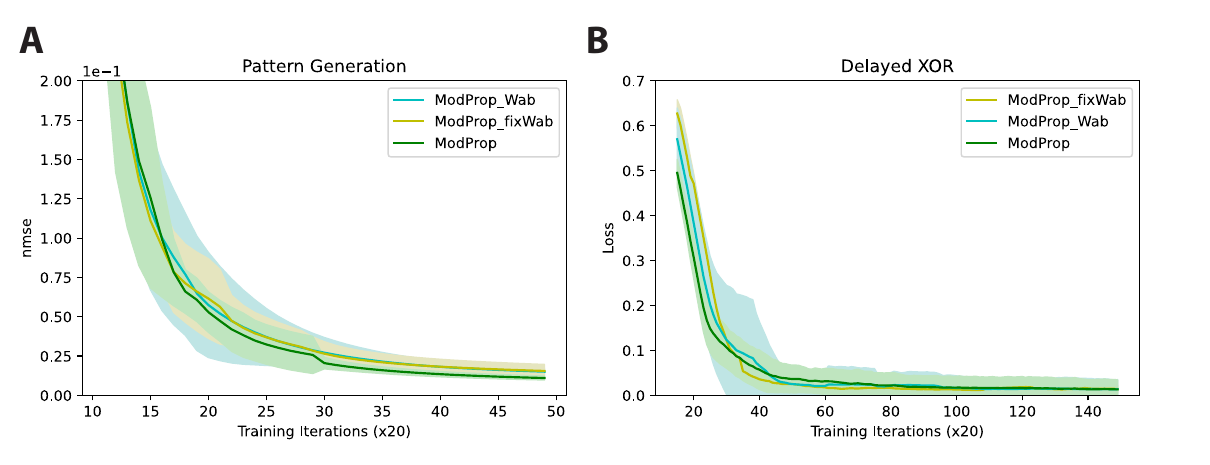}
	\caption{\textbf{Effective learning is achievable with \textbf{fixed} random synapse-type-specific modulatory weights.} Figure~\ref{fig:learningCurves_MDGL++Wab} computes type-specific modulatory weights by averaging forward weight entries in the corresponding pre- and postsynaptic cell group. This assumes that modulatory weights co-adapt with synaptic weights. To what extent they are linked in the brain is unclear. Thus, to test the generality of our learning rule, we re-train using fixed random type-specific modulatory weights and show that leads to negligible performance degradation. Note, sequential MNIST task is not considered in figures that involve synapse-type-specific modulatory weights, as cell types were not considered in that task. Plotting convention follows that of previous figures. }
	\label{fig:LearningCurveMDGL++_FixWab}
\end{figure}

\begin{figure}[h!]
	\centering
	\includegraphics[width=0.99\textwidth]{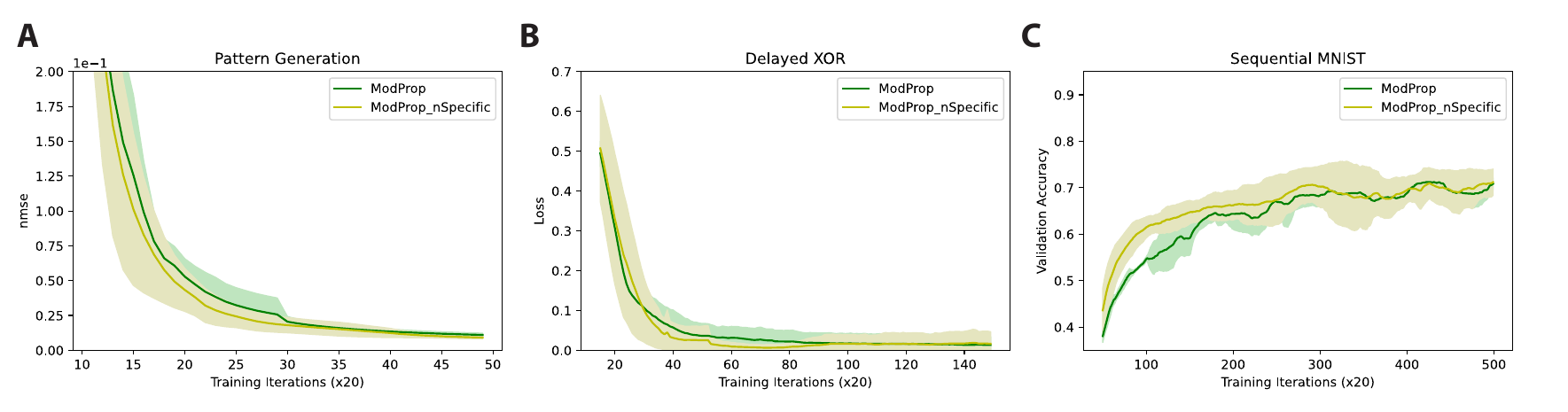}
	\caption{\textbf{Restoring neuron specificity in the activation derivative does not lead to significant improvements.} Here, ModProp\_global is the basic form of ModProp investigated in the main text, where the activation derivative exhibited no spatiotemporal specificity. ModProp\_nSpecific (Eq.~\ref{eqn:MDGL++_nSpec}) takes into account the neuron specificity of the activation derivative and only averages across time steps. This comparison is done for the A) pattern generation task, B)  delayed XOR task and C) sequential MNIST task. Plotting convention follows that of previous figures. }
	\label{fig:learningCurves_MDGL++nSpec}
\end{figure}

\begin{figure}[h!]
	\centering
	\includegraphics[width=0.5\textwidth,trim=9 0 30 0, clip]{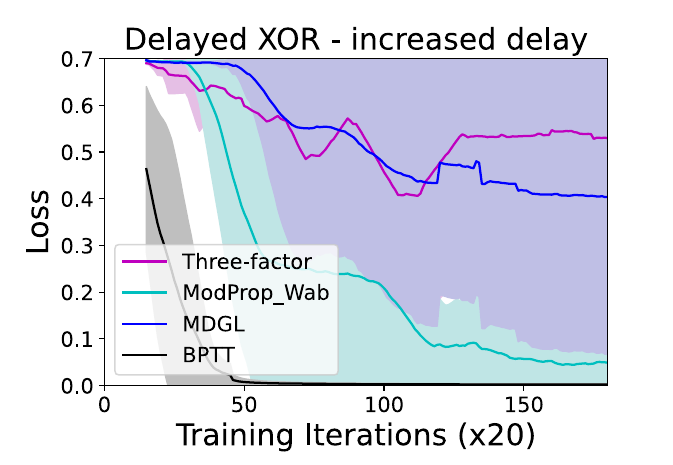}
	\caption{ \textbf{Delayed XOR task with a longer delay period.} We simulate the delayed XOR task with 1.5 times the delay period used in Figure~\ref{fig:learningCurves_MDGL++}B. Although ModProp (with cell-type approximation) still outperforms other bio-plausible learning rules, the performance degrades (compared to ModProp\_Wab in Figure~\ref{fig:learningCurves_MDGL++Wab}B). Moreover, we found all rules (including BPTT) struggle to learn if we increased the delay period to twice of that in Figure~\ref{fig:learningCurves_MDGL++} without changing other task or network parameters (e.g. cue width and intensity). This connects nicely to our discussion point on the limitation of ModProp in addressing very long temporal credit assignment problems in the absence of a long-term memory mechanism. Solid lines/shaded regions: mean/standard deviation of loss curves across five runs. }
	\label{fig:learningCurves_varDly}
\end{figure}

\begin{figure}[h!]
	\centering
	\includegraphics[width=0.99\textwidth,trim=9 0 30 0, clip]{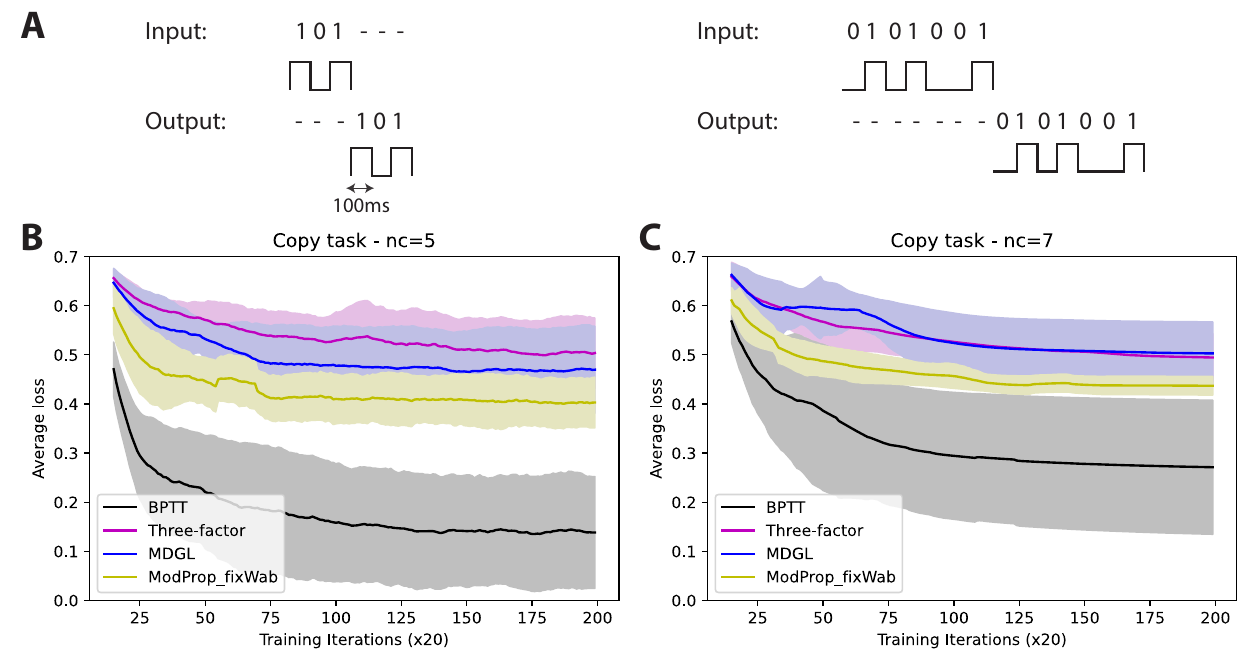}
	\caption{ \textbf{Copy task with fixed random synapse-type-specific modulatory weights.} Sequences of binary cues are presented to an RNN. For each sequence, once the full sequence has been presented, the network should output the original sequence (with the same value and duration) without any further information~\cite{mujika2018approximating}. {\bf A)} Examples of input/output pairs at different sequence lengths. Instead of having each cue lasting just 1 step, we have each cue lasting 100 steps (100ms) to mimic the duration of a quick cue flash in biological settings. Superior performance of ModProp even with fixed and random modulatory weights (compared to other biologically plausible rules) is demonstrated for the copy task with a sequence length of {\bf B)} five cues ($nc=5$) and {\bf C)} seven cues ($nc=7$). Average loss denotes the binary cross entropy loss computed on target and actual output averaged across time steps. Solid lines/shaded regions: mean/standard deviation of loss curves across five runs. }
	\label{fig:learningCurves_copy}
\end{figure}

We discussed how the basic form of ModProp completely neglects any spatiotemporal specificity in the activation derivative. We ask how much performance gain could we get if we lose only temporal specificity, i.e. only average activation derivative across time points. This would see different neurons as having different average activity. To put this more concretely, we approximate the corresponding factor in Eq.~\ref{eqn:pre_nMDGL} as 
\begin{align}
	&\sum_{i_1, \ldots,i_{s-1}} W_{j i_1} h_{i_1,t-1} W_{i_1 i_2} h_{i_2,t-2} \ldots W_{i_{s-1} p} h_{i_{s-1},t-s+1} \cr 
	& \quad \approx W_{j i_1} \overline{h_{i_1}} W_{i_1 i_2} \overline{h_{i_2}} \ldots W_{i_{s-1} p} \overline{h_{i_{s-1}}}  = (\overline{W})^s_{jp}, \label{eqn:MDGL++_nSpec}
\end{align}
where $\overline{W} := W \odot \overline{h}$ with $\overline{h}$ --- a $1-by-N$ vector with each entry corresponding to a neuron-specific mean activation --- broadcasted for the element-wise multiplication with $W$. In other words, this restores spatial specificity and the only approximation being made here is to remove temporal specificity of activation derivative. As a practical note, by the famous AM-GM inequality, the estimation ($\Pi_s h_{s} \approx \overline{h}$) would yield an upper bound of the actual. Thus, we multiply a dampening factor $\mu$ to every $\overline{W}$ for stability, and treat $\mu$ as a hyperparameter. We name this variant of ModProp as "ModProp\_nSpecific", and the most basic form we investigated in the main text as "ModProp\_global". Figure~\ref{fig:learningCurves_MDGL++nSpec} shows that this restoration of neuron-specificity in activation derivative did not lead to significant performance improvement. 

\newpage

\section{Further discussion on related algorithms} \label{scn:related_algs}

For efficient online learning in RNNs, approximations to RTRL have been proposed~\cite{Marschall2019,Mujika2018,Tallec2018,Roth2019,Murray2019,Menick2020,zenke2020spike}. For biological realism (use only local information for local updates) and reduced computational cost, \textbf{e-prop}~\cite{Bellec2020} and \textbf{RFLO}~\cite{Murray2019} proposes severe truncation such that the weight update would only depend on pre- and postsynptic neuron activity as well as a top-down error signal that only tells how a neuron directly contributes to the overall network outcome. \textbf{MDGL}~\cite{liu2021pnas,liu2021solution} proposes a less severe truncation than e-prop and RFLO, but it only addresses the contributions to the task error of neurons that are at most 2 synapses away (note, Ref~\cite{liu2021pnas} can be considered as a special case of our work, where the filter length is constrained to a single tap). ModProp shows a way of removing this significant limitation and enables the communication and calculation of credit from neurons that can be arbitrarily many synapses away. It also experimentally demonstrates the benefit of this key contribution. 

Along the approach of truncations, reference~\cite{Menick2020} proposed the \textbf{SnAp-n} algorithm that allows the user to customize the amount of truncation by deciding on $n$. SnAP-n stores $\frac{\diff s_{j,t}}{\diff w_{pq}}$ (seen in Eq.~\ref{eqn:s_triple}) only for $j$ such that parameter $w_{pq}$ influences the activity of unit $j$ within $n$ time steps. SnAp-1 is closely related to e-prop/RFLO (assuming no autapses). However, starting at $n=2$ (SnAp-2), $\frac{\diff s_{j,t}}{\diff w_{pq}}$ will be stored for every $\{j, p, q\}$ such that $w_{pq}$ influences $j$ in two steps. This would require the storage of $k N^3$ traces, where $k$ is a constant that equals the connection density squared. To our knowledge, there is no evidence on how neural circuits can accommodate such $O(kN^3)$ storage. Therefore, SnAp-n ($n \geq 2$) still poses a significant biological plausibility issue while SnAp-1 reduces to e-prop/RFLO in certain circumstances. 

Moving from temporal truncation, ~\textbf{KeRNL}~\cite{Roth2019} approximates long term dependencies by assuming the dependency to be first order low-pass and learn the parameters using node perturbation. However, the algorithm poses significant implementation and biologically plausibility issues: (1) it uses node perturbation to find the meta-parameters (e.g. the first order decay constant), which is not scalable; (2) meta-parameters are updated per step on the same timescale as synaptic weight update. Their idea of approximating long term dependencies by assuming it follows a certain structure rather than truncating it is what partially inspired our rule. Unlike KeRNL, our approximation can lead to successful learning with fixed meta parameters that are likely updated on the evolutionary timescale in biology (Figure~\ref{fig:LearningCurveMDGL++_FixWab}).

\newpage 

\section{Cost analysis and biological implementation} ~\label{scn:cost}

\subsection{Cost and interpretation for biologically-plausible implementation of ModProp in Eq.~\ref{eq_ModProp}} ~\label{scn:cost_plausible}

Recall for ModProp, the eligibility trace is combined with total modulatory signals detected:
\begin{align}
	\Delta W_{pq} &\propto \text{ET}_{pq} \times \text{TD}_p + \text{ET}_{pq} * \sum_{\alpha\in C} \text{LM}_{\alpha\beta} \cr
	& \text{ET}_{pq} * \sum_{\alpha\in C} \text{LM}_{\alpha\beta} = \sum_{s=1}^S \text{ET}_{pq, t-s} \times \sum_{\alpha\in C} \text{LM}_{\alpha\beta, s} \cr
	\text{LM}_{\alpha\beta, s} &= \text{(affinity }W^s_{\alpha\beta}\text{)} \times \sum_{j\in\alpha} {\underbrace{\strut \text{TD}_j\,\times\,(\text{activity }j)}_{\mathclap{\textstyle \text{\normalsize modulatory signal }j}}}. 
	\label{eq_MDGL++_2}
\end{align}
We see that the eligibility trace (ET) is brought outside of the double summation of local modulatory (LM) signals. A biological interpretation is that secretion of top-down (TD) learning signals can selectively activate a biochemical process at the post-synaptic neuron, which can then act as a temporal filter on the eligibility trace. The number of filter taps for the underlying biochemical process, $S$, determines the number of steps for credit information propagation.

Here is the computational cost breakdown for Remark~\ref{rmk:cost}:
\begin{itemize}
	\item $a_{j,t}=\text{TD}_j\,\times\,(\text{activity }j)$ for all $j=1,..,N$ has $O(N)$ operations. 
	\item $\sum_{j\in \alpha} a_{j,t}$ for $j=1,...,N_{\alpha}$ has $O(N_{\alpha})$ operations (assuming $a_{j,t}$ already available, from the last step), where $N_\alpha$ is the number of cells in type $\alpha$.
	\item $LM_{\beta,s} := \sum_{\alpha \in C} W^s_{\alpha \beta} (\sum_{j\in \alpha} a_{j,t})$ for all $\alpha, \beta=1,...C$ and $s=1,...,S$ has $O(SC^2)$ operations. Note, this step is the modulatory communication step, where type-specific-approximation of weights can reduce the cost. 
	\item $\sum_{s=1}^S \text{ET}_{pq, t-s} \times \text{LM}_{\beta, s}$ for all $p,q=1,...,N$ has $N^2$ element-wise multiplications per $s=1,...,S$, leading to a total of $O(SN^2)$. Since $\beta$ can be determined from $p$, there is no need to loop over $\beta$ in this step.
\end{itemize}
Since the cost of the last item dominates, \textbf{the computational cost scales as $O(SN^2)$}. For storage cost, ModProp stores $e_{pq,t-S}, \dots e_{pq,t}$ for every $(pq)$, leading to \textbf{a storage cost of $O(SN^2)$}. 

\subsection{Cost for biologically-implausible implementation of ModProp} ~\label{scn:cost_implausible}

We prove Proposition~\ref{prop:online} next, where we discussed a potentially biologically-implausible in silico implementation with lower computational and storage costs than the biologically-plausible version above. 
\begin{proof}
	Let us first introduce the following notations:
	\begin{itemize}
		\item $N_\alpha$ denotes the number of cells in type $\alpha$
		\item $[(W^s)_{\alpha \beta}] \in \mathbb{R}^{N_\alpha \times N_\beta}$ is a matrix repeating the value of scalar $(W^s)_{\alpha \beta}$
		\item Thus, $[(W^1)_{\alpha \gamma}] [(W^s)_{\gamma \beta}] = [N_\gamma (W^1)_{\alpha \gamma} (W^s)_{\gamma \beta}]$
		\item $G^t_{\alpha, pq} := \sum_{s=1}^{t} (W^s)_{\alpha\beta}\, \mu^{s-1} \, e_{pq,t-s}$
	\end{itemize}
	
	By properties of block matrix product:
	\begin{align}
		[(W^{s+1})_{\alpha \beta}] &= \sum_\gamma [(W^1)_{\alpha \gamma}] [(W^s)_{\gamma \beta}] = [\sum_\gamma N_\gamma (W^s)_{\alpha \gamma} (W^s)_{\gamma \beta}] \cr
		&\rightarrow (W^{s+1})_{\gamma \beta} = \sum_\gamma N_\gamma (W^1)_{\alpha \gamma} (W^s)_{\gamma \beta}. 
	\end{align}
	
	Now, let's find a recursive expression to calculate $Z^{t+1}_{\alpha, pq}$ online:
	\begin{align}
		G^{t+1}_{\alpha, pq} &= \sum_{s=1}^{t+1} (W^s)_{\alpha\beta}\, \mu^{s-1} \, e_{pq,t+1-s} \cr 
		&= \sum_{s=0}^{t} (W^{s+1})_{\alpha\beta}\, \mu^{s} \, e_{pq,t-s} \cr 
		&= \sum_{s=1}^{t} (W^{s+1})_{\alpha\beta}\, \mu^{s} \, e_{pq,t-s} + (W^1)_{\alpha\beta} e_{pq,t} \cr
		&= \sum_{s=1}^{t} (\sum_\gamma N_\gamma (W^1)_{\alpha \gamma} (W^s)_{\gamma \beta}) \, \mu^{s} \, e_{pq,t-s} + (W^1)_{\alpha\beta} e_{pq,t} \cr
		&= \sum_\gamma N_\gamma (W^1)_{\alpha \gamma} \mu \sum_{s=1}^t (W^s)_{\gamma \beta} \, \mu^{s-1} \, e_{pq,t-s} + (W^1)_{\alpha\beta} e_{pq,t} \cr 
		&= \sum_\gamma N_\gamma (W^1)_{\alpha \gamma} \mu G^t_{\gamma,pq} + (W^1)_{\alpha\beta} e_{pq,t} 
	\end{align}
	
	And the overall update is:
	\begin{align}
		\left. \Delta W_{pq} \right|_{ModProp}
		&= \frac{\partial E}{\partial z_{p,t}} e_{pq,t} + \sum_{\alpha} G^t_{\alpha, pq} \sum_{j \in \alpha} \frac{\partial E}{\partial z_{j,t}} h_{j,t}
	\end{align}
	
	The second term dominates the cost, for which we need to store $G^t_{\alpha, pq}$ for every $\alpha, p, q$. This amounts to $O(CN^2)$ storage cost. To update and attain $G^{t+1}_{\alpha, pq}$, we need $O(C)$ summations and multplications per $G^{t+1}_{\alpha, pq}$, which amounts to $O(C^2N^2)$ computational cost. The final step of combining $G$ and $\sum_{j \in \alpha} \frac{\partial E}{\partial z_{j,t}} h_{j,t}$, requires $O(CN^2)$ computational cost, which does not dominate the cost.
	
	We note that the specific implementation outlined in the proof of Proposition~\ref{prop:online} can significantly reduce the implementation cost, but is likely biologically-implausible, because each synaptic weight update requires the knowledge of all modulatory weights in the network (Appendix~\ref{scn:cost}). Moreover, it reduces the cost compared to RTRL ($O(N^3)$ storage and $O(N^4)$ computational complexity) as well as SnAP-2 ($O(d^2 N^3)$ storage and $O(d^2 N^4)$ computational complexity for connection density $d$)~\cite{Menick2020} significantly if only a few cell types are used. In this work, we used only two cell types ($C=2$) that map onto the two main cell classes: excitatory and inhibitory. However, ModProp is more expensive (by a constant factor) than e-prop, RFLO and MDGL, which all have $O(N^2)$ storage and $O(N^2)$ computational complexity. However, as mentioned, the performance of these rules are limited due to their severe temporal truncation.
	
\end{proof}

\newpage

\section{Unreasonable effectiveness of synapse-type-specific modulatory backpropagation (through time) weights} \label{scn:proofs}

We provide the proof for Theorem~\ref{thm:decE} below:

\begin{proof}
	We first show that $\mathbb{E}[(\epsilon_s)_{ij}]=0$ for all $s\geq 1$. Note $(W^s)_{\alpha \beta}= \sum_{\gamma} N_{\gamma} (W^{s-1})_{\alpha \gamma} W_{\gamma \beta}$ and $(W^s)_{ij} = \sum_{k} (W^{s-1})_{ik} W_{kj}$ can be calculated recursively.
	
	The base case $s=1$ is already given in the condition. Suppose $\mathbb{E}[(\epsilon_s)_{ij}]=0$, for $s+1$:
	\begin{align}
		\mathbb{E}[(\epsilon_{s+1})_{ij}] &= \mathbb{E}[(W^{s+1})_{ij} - (W^{s+1})_{\alpha \beta}] \cr
		&= \mathbb{E}\left[ \sum_k (W^{s})_{ik} W_{kj}\right] - \sum_{\gamma} N_{\gamma} (W^s)_{\alpha \gamma} W_{\gamma \beta} \cr
		&= \mathbb{E}\left[ \sum_\gamma \sum_{k\in\gamma} ((W^{s})_{\alpha \gamma} + (\epsilon_{s})_{ik}) (W_{\gamma \beta}+\epsilon_{kj})\right] - \sum_{\gamma} N_{\gamma} (W^s)_{\alpha \gamma} W_{\gamma \beta} \cr 
		&= \sum_{\gamma} N_{\gamma} (W^s)_{\alpha \gamma} W_{\gamma \beta} + 
		\sum_\gamma \sum_{k\in\gamma} \mathbb{E}[(\epsilon_{s})_{ik}] W_{\gamma \beta} + \sum_\gamma \sum_{k\in\gamma} (W^{s})_{\alpha \gamma} \mathbb{E}[\epsilon_{kj}] \cr
		&\qquad + \sum_\gamma\sum_{k\in\gamma} \mathbb{E}[(\epsilon_{s})_{ik} \epsilon_{kj}] - \sum_{\gamma} N_{\gamma} (W^s)_{\alpha \gamma} W_{\gamma \beta} \cr
		&= \sum_\gamma\sum_{k\in\gamma} \mathbb{E}[(\epsilon_{s})_{ik}] W_{\gamma \beta} + \sum_\gamma\sum_{k\in\gamma} (W^{s})_{\alpha \gamma} \mathbb{E}[\epsilon_{kj}] + \sum_\gamma\sum_{k\in\gamma} \mathbb{E}[(\epsilon_{s})_{ik}] \mathbb{E}[\epsilon_{kj}] \cr
		&= 0. \label{eqn:zero_mean} 
	\end{align}
	
	We now prove the Theorem statement for the scalar output case. The extension to multiple output signals follows identically. Consider the loss decrement after one update, under the (locally) first order loss assumption:
	\begin{align}
		\mathbb{E} \left[ \left. \Delta E \right|_{pq,t} \right] &= -\eta \mathbb{E} \left[ \widehat{\frac{dE}{dW_{pq}}} \frac{dE}{dW_{pq}} \right] \cr 
		&= -\eta \mathbb{E} \Bigg[ (y_t-y_t^*)^2 \bigg[ W_{p}^{\text{OUT}} e_{pq,t} + \sum_{s,\alpha} \sum_{j\in\alpha} W_{j}^{\text{OUT}} (W^s)_{\alpha\beta} e_{pq,t-s} \bigg] \cr & \qquad \times
		\bigg[ W_{p}^{\text{OUT}} e_{pq,t} + \sum_{u,\alpha'} \sum_{j'\in\alpha'} W_{j'}^{\text{OUT}} [(W^u)_{\alpha'\beta}+(\epsilon_u)_{j'p}] e_{pq,t-u} \bigg] \Bigg] \cr
		&= -\mathbb{E} [\Gamma^2_{pq}] -\eta \mathbb{E} \Bigg[ (y_t-y_t^*)^2 \bigg[ W_{p}^{\text{OUT}} e_{pq,t} + \sum_{s,\alpha} \sum_{j\in\alpha} W_{j}^{\text{OUT}} (W^s)_{\alpha\beta} e_{pq,t-s} \bigg] \cr
		& \qquad \times \bigg[ \sum_{u,\alpha'} \sum_{j'\in\alpha'} W_{j'}^{\text{OUT}} (\epsilon_u)_{j'p} e_{pq,t-u} \bigg] \Bigg] \cr
		&= -\mathbb{E}[\Gamma^2_{pq}] -\eta \sum_{u,\alpha'} \sum_{j'\in\alpha'} W_{j'}^{\text{OUT}} W_{p}^{\text{OUT}} \mathbb{E}\left[ (\epsilon_u)_{j'p} (y_t-y_t^*)^2 e_{pq,t} e_{pq,t-u}\right] \cr
		& \qquad -\eta \sum_{s,u,\alpha,\alpha'}\sum_{j\in\alpha, j'\in\alpha'} (W^s)_{\alpha\beta} W_{j}^{\text{OUT}} W_{j'}^{\text{OUT}} \mathbb{E}\left[ (\epsilon_u)_{j'p} (y_t-y_t^*)^2 e_{pq,t-s} e_{pq,t-u} \right]  \cr
		&\overset{(a)}{=} -\mathbb{E}[\Gamma^2_{pq}] -\eta \sum_{u,\alpha'} \sum_{j'\in\alpha'} W_{j'}^{\text{OUT}} W_{p}^{\text{OUT}} \mathbb{E}\left[ (\epsilon_u)_{j'p}\right] \mathbb{E}\left[ (y_t-y_t^*)^2 e_{pq,t} e_{pq,t-u}\right] \cr
		& \qquad -\eta \sum_{s,u,\alpha,\alpha'}\sum_{j\in\alpha, j'\in\alpha'} (W^s)_{\alpha\beta} W_{j}^{\text{OUT}} W_{j'}^{\text{OUT}} \mathbb{E}\left[ (\epsilon_u)_{j'p}\right] \mathbb{E}\left[(y_t-y_t^*)^2 e_{pq,t-s} e_{pq,t-u} \right] \cr
		&\overset{(b)}{=} -\mathbb{E}[\Gamma^2_{pq}] \leq 0, 
	\end{align}
	where $(a)$ follows from the uncorrelatedness condition and $(b)$ follows from the result of \eqref{eqn:zero_mean}. Here, we defined $\Gamma_{pq} := \eta (y_t-y_t^*) \, \left[ W_{p}^{\text{OUT}} e_{pq,t} + \sum_s \sum_\alpha \sum_{j\in\alpha} W_{j}^{\text{OUT}} (W^s)_{\alpha\beta} e_{pq,t-s} \right]$. Then,
	\begin{equation}
		\mathbb{E}[\left. \Delta E \right|_t] = -\eta \mathbb{E}\left[\widehat{\nabla E}^T \nabla E\right] = -\eta \sum_{p,q} \mathbb{E}\left[\widehat{\frac{dE}{dW_{pq}}} \frac{dE}{dW_{pq}}\right] \leq 0.
	\end{equation}
	
	Moreover, if gradient descent is possible for a network $\hat{\mathcal{N}}$ with weight $W_{ij}=W_{\alpha\beta}$, $\forall i\in\alpha, j\in\beta$, then $\mathbb{E}[\sum_{p,q} \Gamma_{pq}]<0$ by definition and $\left. \mathbb{E}[\Delta E \right|_{t}] < 0$.
	
\end{proof}

We note that with the linear RNN assumption in Theorem~\ref{thm:decE}, \textbf{Approximation 1} becomes exact when $\mu=1$ because the activation derivative is a constant $1$ for linear networks. Thus, the proof only only examines the effect of \textbf{Approximation 2} (type-specific feedback weight approximation). Also, the Theorem assumes uncorrelatedness for residual weights $\epsilon$, which may not be the case for networks that are not Erdős–Rényi~\cite{hu2018feedback}. Despite that, ModProp still leads to performance improvement over existing rules for the tasks examined. Nevertheless, it is important to investigate ModProp across a broad range of tasks in the future.  

\begin{figure}[h!]
	\centering
	\includegraphics[width=0.99\textwidth]{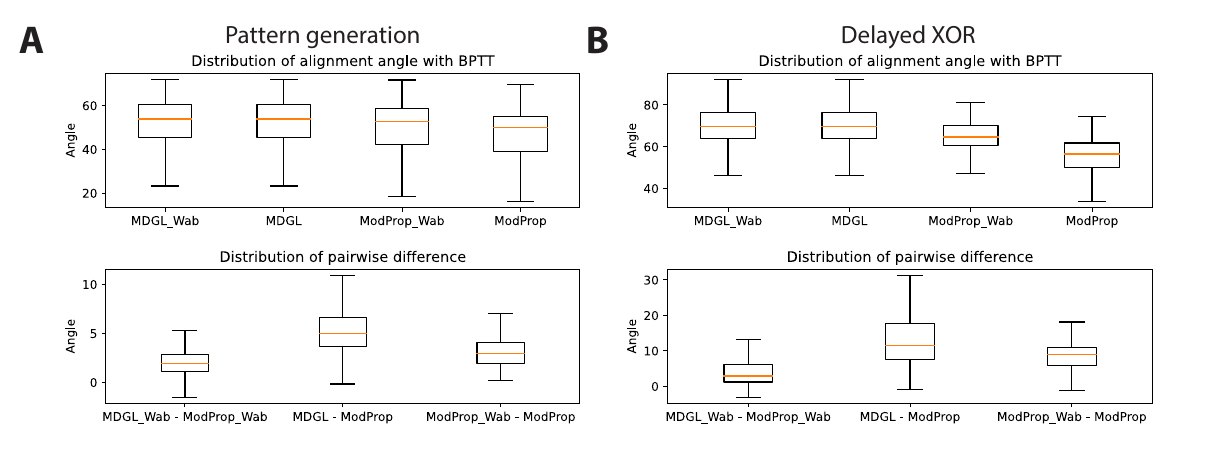}
	\caption{ \textbf{Alignment angle comparison shows that gradients approximated by ModProp (with or without type-specific modulatory weights) are more similar (than MDGL) to the exact gradients}. We quantify the similarity between approximated and exact gradients via the alignment angle, which describes the similarity in the direction of the two update vectors (Appendix~\ref{scn:sim_details}) for various tasks. In all top-panels, the alignment angles between approximate rules and BPTT are less than $90^{\circ}$, which indicate that the approximated gradients are aligned with the exact gradient, despite the high-dimensionality of the update vectors. All bottom panel plots show that ModProp variants achieve smaller alignment angles (hence better alignment) with BPTT than MDGL does. To ensure a fair comparison, we examine the statistics of pairwise differences, so that the point on the loss landscape --- where the comparison is done --- is matched. This is achieved by training the network using BPTT across seven different runs and sampling the approximated gradient once every 50 training iterations. Alignment analysis illustrated here is for recurrent weight gradients, and similar trends are observed for the input weights as well. } 
	\label{fig:plotAlignMDGL++}
\end{figure}

\newpage

\section{Simulation details}
\label{scn:sim_details}
All weight updates were implemented using Adam with default parameters~\cite{Kingma2015}. All Adam learning rates are optimized by picking the best one within $\{5e-5, 1e-4, 2e-4, 5e-4, 1e-3, 2e-3, 5e-3 \}$ for every learning rule and task. For ModProp, the best value of hyperparameter $\mu$ (Eq.~\ref{eqn:approx1}) was picked within $\{0.2, 0.25, 0.3, 0.35, 0.4, 0.45, 0.5 \}$. For every learning rule and task, we removed the worst performing run quantified by area under the learning curve. We note that while input, recurrent and output weights are all being trained, the nonlocality issue (Eq.~\ref{eqn:s_triple}) only applies to training input and recurrent weights. Thus, all approaches update output weights using backpropagation, and approximations apply to training input and recurrent weights. As stated, we repeated runs with different random initialization to quantify uncertainty and weights were initialized similarly as in~\cite{Bellec2018}.

We used alignment angle to quantify the similarity between two vectors. The alignment angle $\theta$ between two vectors, a and b, was computed by $\theta=acos(\|a^Tb\|/\|a\|\|b\|)$. The alignment between two 2D matrices was computed by flattening the matrices into vectors. 

For the pattern generation task, our network consisted of 400 neurons described in Eq.~\ref{eq_rate1}. All neurons had a membrane time constant of $\tau_m = 30\text{ms}$. Input to this network was provided by 50 units each producing a different random Gaussian input. The fixed target signal had a duration of $2000\text{ms}$ and given by the sum of five sinusoids, with fixed frequencies of $0.5\text{Hz}$, $1\text{Hz}$, $2\text{Hz}$, $3\text{Hz}$ and $4\text{Hz}$. For learning, we used mean squared loss function and for visualization, we used normalized mean squared error $\text{NMSE}=\frac{\sum_{k,t} (y^*_{k,t} - y_{k,t})^2}{\sum_{k,t} (y^*_{k,t})^2}$ for zero-mean target output $y^*_{k,t}$. 
For the delayed XOR task, our implementation of the task involved the presentation of two sequential cues, each lasting $100\text{ms}$ and separated by a $700\text{ms}$ delay. There was only one input unit involved and two cue alternatives were presented by setting the input unit to $1$ or $0$, and the unit was set to to $0$ during the delay period. In addition, a Gaussian noise with $\sigma=0.01$ was added to the input. The network was trained to output 1 (resp. 0) at the last time step when the two cues have matching (resp. non-matching) values. Our network consisted of 120 neurons. All neurons had a membrane time constant of $\tau_m = 100\text{ms}$. For learning, we used cross-entropy loss function and the target corresponding to the correct output was given at the end of the trial. A batch size of 32 was used and the gradients were accumulated during those trials additively. 

For the copy task, we presented a input sequence of seven binary cues (taking on the value of $0$ or $1$) on one set of runs and five cues on another. Each cue lasts $100\text{ms}$ to mimic duration of a quick cue flash in biological setting. After the full sequence presentation, the network is tasked to output the same sequence (same value and duration) without further instruction. Our network consisted of 120 neurons. All neurons had a membrane time constant of $\tau_m = 100\text{ms}$. For learning, we used cross-entropy loss function and the target corresponding to the correct output was given at the end of the trial. We used full batch training: a batch size of 8 (resp. 128) was used for the three (resp. seven) cue sequence runs due to 8 (resp. 128) possible permutations. 

For the pixel-by-pixel MNIST task~\cite{lecun1998mnist}, our network consisted of 200 neurons. All neurons had a membrane time constant of $\tau_m = 20\text{ms}$. Input to this network was provided by a single unit that represented the grey-scaled value of a single pixel, with a total of 784 steps and the network prediction was made at the last step. For learning, we used the cross-entropy loss function and the target corresponding to the correct output was given at the end of the trial. A batch size of 256 was used and the gradients were accumulated during those trials additively. 

We used TensorFlow~\cite{abadi2016tensorflow} version 1.14 and based it on top of~\cite{Bellec2018}. \footnote{Our code is available: \url{https://github.com/Helena-Yuhan-Liu/ModProp}.} We performed simulations on a computer server with x2 6-core Intel Xeon E5-2640, 2.5GHz, 32 GB RAM. Regardless of the learning rule, our implementation takes approximately one hour to complete one run of pattern generation or delayed XOR task training (for Figures~\ref{fig:learningCurves_MDGL++} and~\ref{fig:learningCurves_MDGL++Wab}) on the server. 

\end{document}